\documentclass[12pt,a4paper,oneside]{article}

\usepackage{graphicx,amssymb,amsmath}
\usepackage{setspace} % should be before hyperref, otherwise footnote links lead to first page
\usepackage[linkcolor={blue},citecolor={blue},colorlinks=true,linktocpage,urlcolor={blue}]
{hyperref}
\usepackage{cite}
\usepackage[margin=2cm]{geometry}
\usepackage[dvipsnames]{xcolor}
\definecolor{darkblue}{rgb}{0,0,0.54}
\usepackage{tikz}
\usetikzlibrary{decorations.pathmorphing}
\usetikzlibrary{patterns}
\usetikzlibrary{calc}
\usepackage{multirow} % Multi-column and multi-row cells \in table
\usepackage{appendix}
\usepackage{bbm}
\usepackage{import}
\usepackage{tensor}
\usepackage{pdfpages}
\usepackage{cancel} % when want to show that terms cancel, use \bcancel
\usepackage{slashed} % for Dirac slash notation
\usepackage{xparse} % for conditional newcommands (DeclareDocumentCommand)
\usepackage{tabu} % table
\usepackage{adjustbox} % fit to size, use in table
\usepackage{simpler-wick}
\usepackage[bb=boondox]{mathalfa}
\usepackage{dsfont}
\usepackage{subcaption}

\numberwithin{equation}{section} % number equations according to sections
\usepackage{empheq} % emphasizing equations

\graphicspath{{Figures/}} % in order to import figures from a subdirectory

\newcommand{\cO}{\mathcal{O}}

\DeclareMathOperator{\tr}{tr}

\renewcommand{\Re}{\operatorname{Re}}

\newcommand{\mean}{\overline}
\newcommand{\rAdS}{r_{\text{AdS}}}
\newcommand{\rdS}{r_{\text{dS}}}
\newcommand*\dd{\mathop{}\!\mathrm{d}}
\newcommand{\JMS}{\mathbb{J}}
\newcommand{\uint}{\mathfrak{u}} % used for integration variable in an appendix
\newcommand{\uing}{\mathfrak{u'}} % used for integration variable in an appendix
\newcommand{\td}{\mathfrak{t}} % finite time shifts
\newcommand{\ratiocoeff}{g} % notation for a ratio of two integrals
\newcommand{\onecopy}{^{(1)}} % notation for a ratio of two integrals
\newcommand{\projJ}{\mathfrak{J}} % notation for a ratio of two integrals
\usepackage{upgreek}
\newcommand{\mud}{\upmu} % notation for mu in general dimension

\usepackage{titling} % for a tit\le page
\usepackage[affil-it]{authblk} % for the institute etc in the title page

\usepackage{tocloft} % change table of contents font
\title{Towards a microscopic description of de Sitter dynamics}
\author[ ]{Vladimir Narovlansky}
\affil[ ]{School of Natural Sciences, Institute for Advanced Study, Princeton, NJ 08540, USA}
\date{\small \texttt{vladi@ias.edu}}

\makeatletter % use numbers instead of symbols in the thanks part in the title
\let\@fnsymbol\@arabic
\makeatother

\begin{document}

\begin{titlingpage}
    \maketitle
    \begin{abstract}
	Describing dynamics in a gravitational universe with positive cosmological constant, such as de Sitter space, is a conceptually challenging problem. We propose a principle for constructing a quantum system that can potentially be used to study this question. This quantum system describes a heavy object in such a universe interacting with its environment, to which gauge invariant dynamical observables can be anchored. In order to describe gravity with positive cosmological constant, the proposed quantum system needs to agree with all known semiclassical results. We investigate this with a particular microscopic realization constructed using SYK. We first find that correlators match the classical limit of gravity, given by quantum fields in rigid de Sitter space. In particular, the usual UV behavior of quantum fields is surprisingly reproduced by the quantum mechanical system. In order to probe small effects in the gravitational constant, we also consider the intrinsically dynamical out-of-time-order correlators (OTOCs). These correspond to gravitational scattering in the bulk away from the worldline associated with the quantum system. Such OTOCs have highly non-trivial features in de Sitter space, including a Lyapunov exponent which is twice as big as the maximal chaos exponent from the bound on chaos, as well as an unusual behavior of the coefficients in various OTOCs. Interestingly, we find that these features are reproduced by the quantum system.
    \end{abstract}
\end{titlingpage}

\tableofcontents

\section{Introduction} \label{sec:intro}

It is a long standing problem to understand quantum gravitational effects in a universe with positive cosmological constant. One of the main frameworks used for this question is the dS/CFT correspondence \cite{Strominger:2001pn}. In this relation we think about the conformal field theory (CFT) as living at future infinity of de Sitter (dS) space. Such a slice has no time direction which makes it hard to study \emph{dynamical} questions. This is in contrast to spaces such as AdS and the AdS/CFT correspondence \cite{Maldacena:1997re,Gubser:1998bc,Witten:1998qj}. Indeed, in dS/CFT we are mostly concerned with discussing the wavefunction in such a universe using CFT data (see \cite{Maldacena:2002vr}), with no dynamics. More generally, it is challenging to even define observables that make sense for a gravitational system in a closed universe. In particular, there are no such non-trivial local correlation functions, which we usually use in studying dynamics.

\begin{figure}[ht]
    \centering
    \begin{subfigure}[b]{0.45\textwidth}
        \centering
		\begin{tikzpicture}
                \node[fill,inner sep=0pt] (LB) at (0,0) {};
                \node[fill,inner sep=0pt] (LT) at (0,5) {};
                \node[fill,inner sep=0pt] (RB) at (5,0) {};
                \node[fill,inner sep=0pt] (RT) at (5,5) {};
                \draw[thick] (LB)--(LT);
                \draw[line width=2pt, color=blue,->] (RB)--(5,2.5);
                \draw[line width=2pt, color=blue] (RB)--(RT);
                \draw[thick,dashed] (LB)--(RT);
                \draw[thick,dashed] (LT)--(RB);
                \draw[thick] (LT)--(RT);
                \draw[thick] (RB)--(LB);
                \end{tikzpicture}
        \caption{Worldline in de Sitter.}
        \label{fig:worldline_dS_setup}
    \end{subfigure}
    \hfill
    \begin{subfigure}[b]{0.45\textwidth}
        \centering
		\begin{tikzpicture}
                \node[fill,inner sep=0pt] (LB) at (0,0) {};
                \node[fill,inner sep=0pt] (LT) at (0,5) {};
                \node[fill,inner sep=0pt] (RB) at (5,0) {};
                \node[fill,inner sep=0pt] (RT) at (5,5) {};
                \draw[thick] (LB)--(LT);
                \draw[line width=2pt, color=blue,->] (RB)--(5,2.5);
                \draw[line width=2pt, color=blue] (RB)--(RT);
                \draw[thick,dashed] (LB)--(RT);
                \draw[thick,dashed] (LT)--(RB);
                \draw[color=red, decorate,decoration={coil,aspect=0,amplitude=0.5pt}] (5,0.5)--(1,4.5);
                \draw[color=red, decorate,decoration={coil,aspect=0,amplitude=0.5pt}] (1,0.5)--(5,4.5);
                \draw[thick] (LT)--(RT);
                \draw[thick] (RB)--(LB);
                \node[fill,circle,color=red, inner sep=2pt] at (3,2.5) {};
                \end{tikzpicture}
        \caption{Scattering from OTOC.}
        \label{fig:worldline_dS_scattering}
    \end{subfigure}
    \caption{A worldline setup in de Sitter.}
    \label{fig:worldline_dS}
\end{figure}
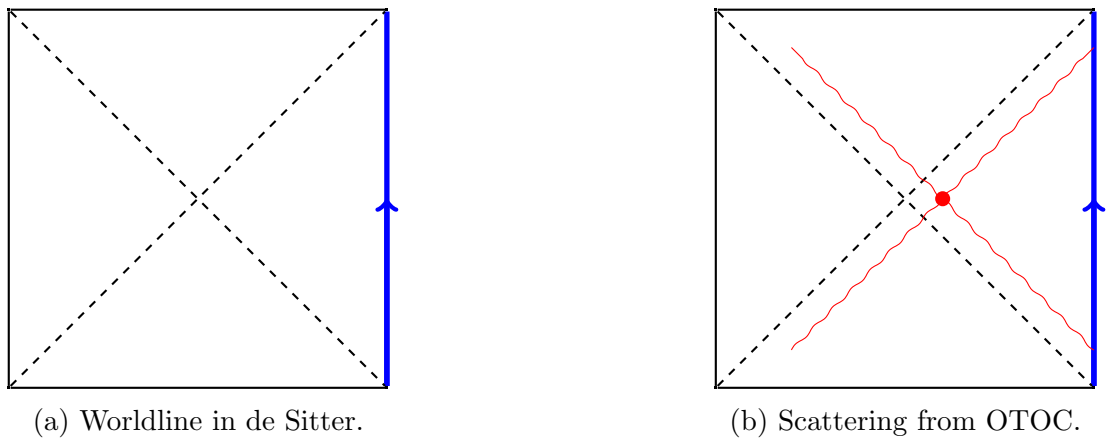

In this paper we will make an attempt to suggest a principle for constructing quantum systems that describe dynamics in dS. In the bulk of the space, we imagine that there is a heavy enough object (in units of inverse de Sitter radius) which will be our anchor for studying well defined questions. Indeed, when we have such an object, it follows a worldline to a good approximation, and in the semiclassical limit where gravitational effects are small it just follows a geodesic, as illustrated in Fig.\ \ref{fig:worldline_dS_setup}. Even when including gravitational effects, such an object has a proper time associated with it, which is gauge invariant and well defined, and we can use it in order to study correlation functions with non-trivial time dependence.

This defines a quantum mechanical system, and the goal of this approach is to give a microscopic description of this quantum system that ideally would agree with everything we know about semiclassical gravity effects.
One possible approach is to construct this system explicitly from what we know about the bulk physics.
Instead, in this paper, we will suggest a simple description of such a quantum mechanical system, building on holographic theories, which we know quite a lot about. This construction a priori knows nothing about bulk physics in a universe with positive cosmological constant. Therefore, the first step is to check whether this construction agrees with known facts about such a universe.

The quantum system is defined by taking two copies of a holographic quantum system and coupling them, by essentially matching states of the same energy. The way we do it might not precisely be what one would usually mean by gauging the difference of their Hamiltonians, but rather gives a rather non-trivial system. The details of the general construction are given in Sec.\ \ref{sec:system}. The reason for eliminating one of the Hamiltonians is intuitively clear from the point of view of the bulk system we described: we would like to remain with a single Hamiltonian that is conjugate to the proper time along the worldline theory.

This construction is a refinement of previous suggestions, such as the interesting discussion of \cite{Anninos:2011af}, and the further study in \cite{Narovlansky:2023lfz}, supplemented by several detailed checks. It also relates to other studies, including among others \cite{Chandrasekaran:2022cip,Kolchmeyer:2024fly}.

One notices pretty quickly that there is a serious difficulty with such an approach. Suppose we are interested in understanding quantum fields in such a bulk theory with positive cosmological constant. The case of a scalar field is reviewed in Sec.\ \ref{sec:scalar_dS}. Such quantum fields have the usual UV singularities when two such fields approach one another, which are there at first approximation where we neglect gravitational effects. However, in a quantum mechanical system in $0+1$ dimensions this is usually not the case.

In order to test this in our proposal, we consider a concrete example discussed in Sec.\ \ref{sec:micro_observables}. As a concrete system we take double-scaled SYK (DSSYK). We use this system mainly because it satisfies the desired holographic properties we need in a particular regime, and it is a UV complete and well defined quantum system. It is our goal to have a microscopic system and that is why it serves this purpose well. It is also convenient as many exact results are known about it. We stress, however, that the main features we use are related to its holographic nature and are not particular to it. In particular, for the semiclassical analysis one could use alternative systems such as the Schwarzian theory, which is not defined microscopically in the same fashion as SYK is.
Moreover, our gravitational setup (demonstrated in Fig.\ \ref{fig:worldline_dS_setup}) is applicable in any bulk dimension, but for simplicity and concreteness we consider mostly three spacetime dimensions.
The only part where we comment on general dimension is Sec.\ \ref{sec:4pf_expansion}.

We find that the UV behavior mentioned before is explicitly obeyed in the constructed two-copy quantum system. This is one of our main results. Moreover, the semiclassical limit of correlation functions agrees with correlators of quantum fields in de Sitter when the gravitational constant is negligible.

In this discussion, in fact, we work in leading order in the gravitational constant $G_N$, and so it does not appear at all. This is the case for quantum fields in rigid de Sitter space. Our goal is to test the gravitational semiclassical limit, and so we would like to go to the next order. One may wonder what would be the next good observable to discuss in the worldline setup we study that is sensitive to $G_N$. We naturally propose to consider out-of-time-order (OTO) correlation functions, and in particular the 4-point out-of-time-order correlator (OTOC). This is indeed an intrinsically dynamical object, just as desired, which is hard to study in quantum gravitational de Sitter space with the common approaches.

Just as black holes, even empty de Sitter space has interesting highly chaotic \cite{Susskind:2011ap} behavior. The origin of this is essentially because dS corresponds to an expanding universe so that objects in such a space are surrounded by a cosmological horizon beyond which they cannot see, because of the fast expansion. However, interestingly, dS has important differences compared to black holes and AdS space, as we review in Sec.\ \ref{sec:review_gravity_OTOC}. One way to characterize chaos is through the OTOC and the Lyapunov exponent. In the gravitational system it can be understood in terms of shockwave geometries in the corresponding space. Such shockwaves in fact behave differently in AdS and in dS. While in AdS they cause time delay, in dS their effect is time advance. This is a dramatic difference, and it manifests itself by making it possible to send a signal between two opposite poles of the space, which is otherwise impossible.

Importantly, the worldline construction makes it possible to study such OTOCs which can live completely on the worldline, and still probe these interesting features. Indeed, one way to think about any 4-point OTOC is through a scattering process in the bulk \cite{Shenker:2013pqa,Roberts:2014isa,Shenker:2014cwa}. So even though our quantum system is really localized to the line, it actually probes a gravitational scattering of particles that create shocks and collide far from the line, as shown in Fig.\ \ref{fig:worldline_dS_scattering}. This worldline OTOC exhibits very interesting properties \cite{Aalsma:2020aib} including a doubled Lyapunov exponent as defined through the OTOC, that is, twice the maximal exponent. These features may indicate that a quantum description of such chaotic behavior should be highly non-trivial.

In Sec.\ \ref{sec:OTOC_4pf} we move on to consider the OTOC in our proposed quantum system, using the particular realization of SYK. Interestingly, we find that both the doubled exponent, as well as the non-trivial behavior of the coefficient in front of it (both differing from the AdS case) are present in the microscopic system we defined, just as in de Sitter.
This is also one of our main results. We describe the tension with the bound on chaos and the features of our system that allow it. 

We make a few concluding remarks relating to other works in Sec.\ \ref{sec:remarks} and some calculations are deferred to appendices.

\section{The system} \label{sec:system}

Given a holographic quantum system with Hamiltonian $H$ we construct a new system with its own partition function and correlation functions. The new system is constructed using two copies of the initial quantum system, labeled (1) and (2), with Hamiltonians $H^{(1)} $ and $H^{(2)} $. Intuitively we would like to gauge $H^{(1)} -H^{(2)} $. The reason to do that is in order to remain with a single time direction in the resulting system, which is the variable conjugate to $\frac{1}{2} \big(H^{(1)} +H^{(2)} \big)$. The resulting system can be thought as being slightly different from what one might naively mean when gauging $H^{(1)} -H^{(2)} $, since by the latter one could mean a system that is essentially isomorphic to a single copy of the original system, with the same partition function. This is not the case for the system we construct.

There are two ways to define the system. One explicit way to describe it is by specifying its ingredients. We can phrase it in terms of the tensor product of the two copies of the system.
\begin{itemize}
\item The states in the doubled system are given by the diagonal states. In particular, if $|E\rangle $ are energy eigenstates in the original system, states in the doubled system are spanned by $|E\rangle _1 |E\rangle _2$ where the subscript describes the corresponding tensor product factor.
\item Operators at a given time $t$ are spanned by
\begin{equation} \label{eq:ops_doubled_system}
\frac{1}{2\pi} \int_{-\infty } ^{\infty }  dt' \, \cO_1 ^{(1)} (t') \cO_2 ^{(2)} (t-t')
\end{equation}
where $\cO _{1,2} $ are two possibly different operators, and the superscript indicates on which copy the operator acts. The time variable $t$ in these operators is indeed conjugate to $\frac{1}{2} \big(H^{(1)} +H^{(2)} \big)$, and these operators commute with $H^{(1)} -H^{(2)} $.

\end{itemize}

In correlation functions, because of separate time translation invariance in each system, nothing will depend on the sum of the integrated time variables ($t'$ above), resulting in an extra $\delta (0)$; this is a common factor to all unnormalized correlators which we can scale away in the normalized correlation functions.

One could consider correlation functions in any of the states described above, and intermediate states are automatically in the same class. Note that in a semiclassical system, preparing a thermal state coupling to $\frac{1}{2} \big( H^{(1)} +H^{(2)} \big)$ would naturally concentrate at the same energies on both sides, giving effectively a state in the defined system.

An alternative specification of the system is again to start with the tensor product of two copies of the system and use the following
\begin{itemize}
\item States in the theory are obtained by using projection operators to enforce the equal energy constraint $\Pi = \int \frac{d\eta}{2\pi } \, e^{i \eta ( H^{(1)} -H^{(2)} )} $.
\item Operators are again spanned by \eqref{eq:ops_doubled_system}.\footnote{More precisely, these should be appropriately normalized, and can be written as $ \lim _{\tau \to \infty } \frac{1}{2\tau } \int _{-\tau } ^{\tau } dt' \, \cO _1^{(1)} (t') \cO _2^{(2)} (t-t')$. In other words, since in this approach the states (including intermediate states) are already invariant under the difference of the Hamiltonians due to the projection, the $t'$ translation does nothing and we should divide by the volume of the orbit.}
\end{itemize}

In fact, with this definition there is a natural ambiguity. Dimensional analysis would tell us that the projector $\Pi $ should come with a constant $\projJ$ with dimensions of energy. In the rest of the discussion we set $\projJ=1$ which can easily be restored, and mention this parameter when needed.

\section{2-point function in de Sitter} \label{sec:scalar_dS}

We review first the case where we can ignore gravitational interactions, leaving us with matter living on rigid de Sitter space. We concentrate on the case of a free massive scalar (see, e.g., \cite{Bousso:2001mw}), where the basic ingredient in the quantum theory of such a field is its propagator, or the 2-point function.

$d$-dimensional de Sitter space can be viewed as a hyperboloid in $d+1$ dimensions. Indeed, in terms of the embedding coordinates $X^A$ it is given by
\begin{equation}
\eta _{AB} X^A X^B=\rdS^2,\qquad \eta _{AB} =\text{diag}(-1,+1,\cdots ,+1).
\end{equation}
Let us consider the Wightman 2-point function
\begin{equation}
G(X,Y)=\langle \phi (X)\phi (Y)\rangle .
\end{equation}
For a state invariant under the $SO(1,d)$, the 2-point function $G(X,Y)$ can only be a function of $X \cdot Y$, which we can parametrize using $P$ defined by $X \cdot Y = \rdS^2 P$. The 2-point function satisfies the Klein-Gordon equation
\begin{equation}
(\nabla ^2 - m^2)G=0
\end{equation}
for a scalar of mass $m$. The Laplacian can be expressed in terms of the embedding coordinates as
\begin{equation}
\rdS^2 \nabla ^2=X^2 \partial _X^2-d X \cdot \partial _X-X^A X^B\partial _{X^A} \partial _{X^B} .
\end{equation}
So in terms of $P$ the Klein-Gordon equation is then
\begin{equation} \label{eq:KG_using_P}
(1-P^2)G''(P)-dP G'(P)-m^2 \rdS^2 G(P)=0.
\end{equation}
The solution of this equation describing the usual Bunch-Davies vacuum is \footnote{In the following the 2-point function behaves for small spacelike geodesic distance $x$ as $\frac{\Gamma \left( \frac{d-2}{2} \right) }{4\pi ^{d/2} x^{d-2} }$ as needed, fixing the coefficient.}
\begin{equation} \label{eq:2pf_dS_2F1}
G(P) = \frac{\Gamma (2\Delta _+)\Gamma (2\Delta _-)}{\rdS^{d-2}(4\pi )^{d/2} \Gamma \left( \frac{d}{2} \right) } {}_2F_1\left( 2\Delta _+,2\Delta _-,\frac{d}{2} ,\frac{1+P}{2} \right) 
\end{equation}
where $\Delta _{\pm } $ are solutions to \footnote{Note that $\Delta $ is related to the more conventional $h$ used in dS/CFT by $h=2\Delta $.}
\begin{equation}
4\Delta \left( \frac{d-1}{2} -\Delta \right) =m^2 \rdS^2.
\end{equation}

We can also express the invariant distance in terms of the geodesic distance between the two points. For spacelike separation $P$ is cosine of the separating angle, while for timelike separation we can write $P= \cosh(t /\rdS)$ where $t $ is the geodesic distance. In terms of the geodesic distance the solution often simplifies. Using this variable, equation \eqref{eq:KG_using_P} becomes
\begin{equation}
\left( \partial _{t } ^2+(d-1) \rdS^{-1} \coth \Big(\frac{t }{\rdS} \Big) \partial _{t } +m^2\right) G(t )=0 .
\end{equation}
For instance, for $d=3$ and working in units where $\rdS=1$, this equation has two independent solutions taking a simple form
\begin{equation} \label{eq:2pf_3d_time_domain}
G_1(t ) = \frac{\sinh\big(\mu (t +i\pi )\big)}{\sinh(t )} ,\qquad 
G_2(t )=\frac{\sinh(\mu t )}{\sinh(t )} ,\qquad \mu ^2=1-m^2
\end{equation}
(one could in principle shift $t $ in the numerator of the first solution by different values).

\section{Observables in the microscopic system} \label{sec:micro_observables}

There is an important difference between the two solutions appearing in \eqref{eq:2pf_3d_time_domain}. The first is singular when $t \to 0$ while the second is not. Indeed, the first solution is the one corresponding to the usual Bunch-Davies vacuum. It has the familiar short distance (UV) singularity of quantum fields. The second solution is related to the so-called $\alpha $ vacua, and has a singularity for antipodal points.

Naturally one is interested in the Bunch-Davies vacuum. This raises a significant difficulty, since in a quantum system we would not expect any such short distance singularity; indeed, for instance in a fermionic quantum mechanical system, there is no divergence when two fermions approach each other. This is in tension with the desire to describe de Sitter space. Indeed, this raised some complications in the past.\footnote{In the discussion of \cite{Narovlansky:2023lfz}, two operators were used instead of one, making a KMS condition more tricky.}

The first test of a microscopic system which is supposed to describe de Sitter space is whether it agrees with the results in de Sitter in the semiclassical limit when gravity effects are weak.

\subsection{Microscopic system}

In order to comply with the proposal, we would like a microscopic system that has a semiclassical limit where it is holographic. This can be achieved with double-scaled SYK. This is indeed a microscopic system, constructed using interacting fermions, with a Hamiltonian, so it is well behaved at any energy scale.

More concretely, the system is constructed by considering $N$ Majorana fermions $\psi _i$, with $i=1,\cdots ,N$ and a Hamiltonian
\begin{equation}
H = i^{p/2} \sum _{i_1<\cdots <i_p} J_{i_1 \cdots i_p} \psi _{i_1} \cdots \psi _{i_p} .
\end{equation}
The Hamiltonian just means that every $p$ out of the $N$ fermions interact with each other. One could consider this Hamiltonian for any natural even $p$, but the double-scaling limit is when both $N,p \to \infty $ while keeping
\begin{equation}
\lambda =\frac{2p^2}{N} 
\end{equation}
fixed. $\lambda $ can be thought of as the coupling constant of the theory. An alternative notation that is often used is
\begin{equation}
q=e^{-\lambda } .
\end{equation}
The model is usually considered as a disordered theory so that the coefficients $J_{i_1 \cdots i_p} $ are random couplings, which for simplicity can be taken from independent Gaussian distributions. We denote averaging with respect to this distribution by an overbar. The average of the coefficients is $ \mean{J_{i_1 \cdots i_p} }=0$ and a common normalization for the variance is
\begin{equation}
\mean{J^2_{i_1 \cdots i_p} } = \frac{N \JMS^2}{2p^2 \binom{N}{p} } .
\end{equation}
We wrote this normalization since it is a common choice.\footnote{For our purposes, an alternative normalization of $ \mean{J^2_{i_1 \cdots i_p} } = \frac{N^3 J^2}{8p^6 \binom{N}{p} } $ might be appropriate.}

The operators in this theory are spanned simply by strings of fermions. It is convenient to parametrize how ``heavy'' the operators are by a parameter $\Delta $. In a low energy holographic regime of the theory, $\Delta $ indeed plays the role of scaling dimension. More precisely, the operators can be written as
\begin{equation} \label{eq:single_side_operators}
\cO^{(1)}  _{\Delta } = i^{p\Delta /2} \sum _{i_1 < \cdots < i_{p\Delta } } M_{i_1 \cdots i_{p\Delta } } \psi _{i_1} \cdots \psi_{i_{p\Delta } }
\end{equation}
consisting of $p\Delta $ fermions.\footnote{The superscript for the operators is used to indicate that these are operators in a single copy of double-scaled SYK.} In the double-scaling limit $\Delta $ becomes a real and positive continuous parameter. For more details on the operators, see \cite{Berkooz:2018jqr}.

The energies in the theory go over a compact range when $\lambda $ is finite. It is common to parametrize them by real variables $s$ as
\begin{equation}
E(s)=-\frac{2 \JMS \cos(\lambda s)}{\sqrt{\lambda (1-q)}} ,\qquad s \in \left[ 0,\frac{\pi }{\lambda } \right] .
\end{equation}
We will mostly be interested in the semiclassical limit. As mentioned, $\lambda $ plays the role of a coupling constant, and so we restrict to the limit $\lambda  \to 0$ (or $q \to 1$) \cite{Goel:2023svz,Mukhametzhanov:2023tcg,Okuyama:2023bch}. The simplest holographic description is when we follow the $\lambda  \to 0$ limit by a limit of large energies above the ground state $\frac{E-E_0}{\JMS \lambda } \gg 1$.\footnote{Here $s$ does not scale with $\lambda $.} We will refer to this combined limit as the \emph{holographic-semiclassical limit}.

Next we should consider two copies of this system. One should specify what is the relation between the probability distribution of couplings between the two copies, however, it should be kept in mind that this system is self-averaging and the details of the couplings are not very important in the double-scaling limit. In particular, typical values of the couplings will give the same results as averaged values.

The operators in the doubled system consist of operators of the form
\begin{equation} \label{eq:doubled_generic_ops}
\int dt' \, \cO ^{(1)} _{\Delta ^{(1)} } (t') \cO ^{(2)} _{\Delta ^{(2)} } (t-t')
\end{equation}
with independent $\Delta ^{(1)} $ and $\Delta ^{(2)} $. We will demonstrate our calculations for the case of three dimensional de Sitter space and a massive scalar, in which case the relevant operators are
\begin{equation} \label{eq:doubled_3d_ops}
\cO _{\Delta } (t) = \frac{1}{2\pi } \int dt' \, \cO ^{(1)} _{1-\Delta } (t') \cO ^{(2)} _{\Delta } (t-t')
\end{equation}
parametrized by a single number $\Delta $.
Note that the doubled system is a construction used to give us the desired system, and once we form it we should really think about it as a single quantum mechanical system. In particular, this is why we denoted the two copies by (1) and (2) rather than `left' and `right'.

We should also specify the state in which we propose to calculate the correlation functions. As mentioned, the states in the doubled system are of the form $|E\rangle _1 |E\rangle _2$. One would naively propose to use thermal states, in analogy to the description of black holes in AdS/CFT.
This is indeed what we do, considering thermal states with temperature $T$.\footnote{In particular, we do not consider the maximal entropy state, as was done in \cite{Narovlansky:2023lfz}.} We would also like to point out that in the semiclassical limit, a thermal state is peaked at a particular energy, or a particular large value of $s=s_0$. Therefore, we find it convenient to instead consider correlation functions in this state, and the two are similar up to an overall constant
\begin{equation} \label{eq:thermal_simpler}
\langle \cdots \rangle _T \to \langle s_0|_1 \langle s_0|_2 \cdots |s_0\rangle _1 |s_0\rangle _2.
\end{equation}
From here on, when considering correlation functions we mean the latter correlators.

\subsection{Correlation functions}

In double-scaled SYK there is a convenient diagrammatic form \cite{Berkooz:2018jqr} (analogously to \cite{Mertens:2017mtv}) where we assign rules to elements of a diagram that give the results of correlation functions at all energy scales. We can then use these results in order to infer analogous rules to the constructed doubled system.

The basic idea is to draw a circle, which can be thought of as the thermal circle. We then mark points on the circle, corresponding to insertions of operators. We should go over all pairings of operators of the same flavor and connect them with chords.

The rules assigned to the elements of the diagram in the doubled system are
\begin{itemize}
\item 
Between any two operators there is a segment along the circle where an energy eigenstate propagates with the time evolution
\begin{equation}
\tikz[baseline=0ex]{
	\node[circle,fill,inner sep=1pt] (a) at (0,0) {};
	\node[circle,fill,inner sep=1pt] (b) at (1.5,0) {};
	\draw[thick] (a) to [bend left=60] node[pos=0.5,above]{{\small $s$}} (b);
	\node at (-0.3,0) {{\small $t_2$}};
	\node at (1.8,0) {{\small $t_1$}};
} = e^{-iE(s)(t_2-t_1)}
\end{equation}
where we used Lorentzian time; alternatively, one could use Euclidean time with the usual Euclidean evolution.
\item Integrate over the energies that can propagate in the segments with the measure
\begin{equation}
\frac{\lambda ^{3/2} (1-q)^{9/2} (q;q)_{\infty } ^6}{8\pi ^2 \JMS \sin(\lambda s) \Gamma _q(\pm 2is)^2} ds .
\end{equation}
As mentioned, this generally comes with a factor of $\projJ$.

\item A node on the circle corresponding to an operator is assigned a matrix element $\langle E_2|\cO _{\Delta } |E_1\rangle $. Explicitly, we have
\begin{equation}
\tikz[baseline=0ex]{
	\draw[thick] (1,0) arc (0:60:1) node[pos=0.8,right]{{\small $\, s_2$}};
	\draw[thick] (1,0) arc (360:300:1) node[pos=0.8,right]{{\small $\, s_1$}};
	\draw[thick] (-0.2,0)--(1,0) node[pos=0.2,above]{{\small $\Delta$}};
	\node[circle,fill,inner sep=1pt] at (1,0) {};
} = \frac{(1-q)^{\Delta ^{(1)} +\Delta ^{(2)} -3} }{(q;q)_{\infty } ^3} \sqrt{\frac{\Gamma _q\big(\Delta ^{(1)} +i(\pm s_1 \pm s_2)\big)\Gamma _q\big(\Delta ^{(2)} +i(\pm s_1 \pm s_2)\big)}{\Gamma _q(2\Delta ^{(1)} )\Gamma _q(2\Delta ^{(2)} )} } .
\end{equation}
Here we allowed general $\Delta ^{(1)} $ and $\Delta ^{(2)} $ corresponding to the operators \eqref{eq:doubled_generic_ops}, while for the operators \eqref{eq:doubled_3d_ops} we should use $1-\Delta $ and $\Delta $.

\end{itemize}

First we can start with the analogue of the partition function. In diagrams, this is simply
\begin{equation}
\tikz[baseline=0ex]{
	\draw[thick] (0,0) arc (0:360:1) node[pos=0.25,above]{{\small $s$}};
}
\end{equation}
while we calculate
\begin{equation}
\tr \left[ e^{-\beta  \frac{1}{2} (H^{(1)} +H^{(2)} )} \Pi \right] 
\end{equation}
with the trace in the doubled Hilbert space (and $\beta =1/T$).

This is simple to evaluate using the diagrammatic rules, which in the semiclassical-holographic regime gives
\begin{equation}
Z = \frac{\lambda ^5 (q;q)_{\infty } ^6}{8\pi ^4 \JMS} \int ds \, s \, e^{4\pi s} e^{\frac{2\beta \JMS \cos(\lambda s)}{\lambda } } .
\end{equation}
We have written the approximation of the partition function where the energies and temperature are large. Note that while the effective density of states in general is non-trivial (in the energy basis it is proportional to the square of the single-copy density), the spectrum in double-scaled SYK is compact so the full integrals always converge, and moreover the density of states is often normalized in a way that is independent of the actual number of states.\footnote{In the doubled system the total number of states is affected by the ambiguity mentioned (related to $\projJ$), which is part of the definition of the theory.}

The 2-point function is represented by
\begin{equation} \label{eq:2pf_diagram}
\tikz[baseline=0ex]{
	\draw[thick] (1,0) arc (0:360:1) node[circle,fill,inner sep=1pt,pos=0]{} (a) node[circle,fill,inner sep=1pt,pos=0.5]{} (b) node[pos=0.25,above]{{\small $s_0+\omega $}} node[pos=0.75,below]{{\small $s_0$}};
	\draw[thick] (-1,0)--(1,0) node[pos=0.5,above]{{\small $\Delta $}};
} 
\end{equation}
Here we use the simpler representation \eqref{eq:thermal_simpler} of the thermal state with temperature $T$, so that the arc corresponding to $s_0 \approx \frac{2\pi T}{\lambda \JMS} \gg 1$ is not assigned any value. The fluctuations of $\omega $ in the propagating state are of order 1. Using the diagrammatic rules above, the value assigned to this diagram is simple enough
\begin{equation} \label{eq:micro_2pf_3d}
\langle \cO (t) \cO (0)\rangle = \frac{\lambda ^2 s_0^2}{4\pi ^3T \, \Gamma (2\Delta )\Gamma (2-2\Delta )} \int_{-\infty } ^{\infty }  d\omega \, e^{(2\pi -4\pi iTt)\omega } \Gamma (\Delta  \pm i\omega )\Gamma (1-\Delta  \pm i\omega ).
\end{equation}
Note that the power of $\lambda $ is really $2\Delta ^{(1)}+2\Delta ^{(2)} $ more generally and in particular can be associated to the operator. More details can be found in App.\ \ref{sec:app_2pt_details}. Another way to write it is
\begin{equation} \label{eq:micro_2pf_3d_2}
\begin{split}
\langle \cO (t) \cO (0)\rangle &= \frac{\lambda ^2 s_0^2}{4\pi T \, \Gamma (2\Delta )\Gamma (2-2\Delta )} \int _{-\infty } ^{\infty } d\omega  \frac{e^{(2\pi -4\pi iTt)\omega } }{\cosh\left( \pi \left( \omega  \pm \frac{i\mu }{2} \right) \right) }  = \\
& = \frac{\lambda ^2s_0^2}{2\pi^2 T(2\Delta -1)} \frac{\sinh\left( \mu (2\pi Tt+i\pi )\right) }{\sinh( 2\pi Tt-i\epsilon )} 
\end{split}
\end{equation}
where the $i\epsilon $ is the $i\epsilon $-prescription, and throughout we parameterize $\Delta $'s by an alternative $\mu $, where these two are related by
\begin{equation}
\mu =1-2\Delta .
\end{equation}
We see that this indeed matches the 2-point function in the Bunch-Davies vacuum (see \eqref{eq:2pf_3d_time_domain}) in three dimensions. In \eqref{eq:2pf_3d_time_domain} we used units where the de Sitter temperature is $T_{\text{dS}} =\frac{1}{2\pi } $, so we get the same result, where importantly we see that the microscopic (SYK) temperature agrees with the de Sitter temperature
\begin{equation}
T =  T_{\text{dS}} .
\end{equation}
From the microscopic point of view, one might not be surprised since it obeys the KMS condition. Indeed we should verify that the 2-point function satisfies $G(t-i\beta ) = G(-t)$. The ratio of $\sinh$'s in \eqref{eq:micro_2pf_3d_2} goes under $t \to t-i\beta $ to
\begin{equation}
\frac{\sinh\left( \mu (2\pi Tt-i\pi )\right) }{\sinh(2\pi Tt)}  = \frac{\sinh\left( \mu (-2\pi Tt+i\pi )\right) }{\sinh(-2\pi Tt)} 
\end{equation}
and so indeed satisfies KMS.

This result realizes the worldline setup in de Sitter described in Sec.\ \ref{sec:intro}. In the semiclassical limit, the heavy object follows a geodesic, and the time in the microscopic system maps to the proper time along the worldline in the bulk.

There are also two possible uncrossed 4-point functions. The first is
\begin{equation} \label{eq:4pf_uncrossed_diagram_1}
\tikz[baseline=0ex]{
	\draw[thick] (1.5,0) arc (0:360:1.5)
		node[pos=0,right]{{\small $s_0+\omega_1 $}}
		node[pos=0.25,above]{{\small $s_0 $}}
		node[pos=0.5,left]{{\small $s_0+\omega_2 $}}
		node[pos=0.75,below]{{\small $s_0 $}};
	\node[circle,fill,inner sep=1pt] (a) at ({1.5*cos(45)},{-1.5*sin(45)}) {};
	\node[right] at ({1.5*cos(45)},{-1.5*sin(45)}) {{\small $\, t_1$}};
	\node[circle,fill,inner sep=1pt] (b) at ({1.5*cos(45)},{1.5*sin(45)}) {};
	\node[right] at ({1.5*cos(45)},{1.5*sin(45)}) {{\small $\, t_2$}};
	\node[circle,fill,inner sep=1pt] (c) at ({-1.5*cos(45)},{1.5*sin(45)}) {};
	\node[left] at ({-1.5*cos(45)},{1.5*sin(45)}) {{\small $t_3$}};
	\node[circle,fill,inner sep=1pt] (d) at ({-1.5*cos(45)},{-1.5*sin(45)}) {};
	\node[left] at ({-1.5*cos(45)},{-1.5*sin(45)}) {{\small $t_4$}};
	\draw[thick] (a) to [bend left=60] node[pos=0.5,right]{{\footnotesize $\Delta_1$}} (b);
	\draw[thick] (c) to [bend left=60] node[pos=0.5,left]{{\footnotesize $\Delta_2$}} (d);
} 
\end{equation}
We note that in such diagrams, there is an additional constraint, part of the diagrammatic rules, that the energies in boundary regions that are connected through the bulk of the diagram have the same energy \cite{Berkooz:2018jqr}, so the energy can be thought of as being attributed to bulk regions. In this diagram there are two energies we should integrate over, associated to $\omega _1$ and $\omega _2$. When looking at the diagrammatic rules applied to this diagram, it is clear that this diagram is the same as a product of 2-point functions, since the rules are only sensitive to what happens on both sides of a connecting chord. So each chord looks the same as the one in \eqref{eq:2pf_diagram}. For completeness we give the full expression in App.\ \ref{sec:app_uncrossed_4pt_details}. So we get
\begin{equation}
\langle \cO _2(t_4) \cO _2(t_3) \cO _1(t_2) \cO _1(t_1) \rangle = \langle \cO _2(t_4) \cO _2(t_3)  \rangle \langle \cO _1(t_2) \cO _1(t_1) \rangle .
\end{equation}
This is generally true in the state \eqref{eq:thermal_simpler}.

The other uncrossed 4-point function is
\begin{equation} \label{eq:4pf_uncrossed_diagram_2}
\tikz[baseline=0ex]{
	\draw[thick] (1.5,0) arc (0:360:1.5)
		node[pos=0,right]{{\small $s_0+\omega_1 $}}
		node[pos=0.25,above]{{\small $s_0 +\omega_1+\omega_2 $}}
		node[pos=0.5,left]{{\small $s_0+\omega_1 $}}
		node[pos=0.75,below]{{\small $s_0 $}};
	\node[circle,fill,inner sep=1pt] (a) at ({1.5*cos(45)},{-1.5*sin(45)}) {};
	\node[right] at ({1.5*cos(45)},{-1.5*sin(45)}) {{\small $\, t_1$}};
	\node[circle,fill,inner sep=1pt] (b) at ({1.5*cos(45)},{1.5*sin(45)}) {};
	\node[right] at ({1.5*cos(45)},{1.5*sin(45)}) {{\small $\, t_2$}};
	\node[circle,fill,inner sep=1pt] (c) at ({-1.5*cos(45)},{1.5*sin(45)}) {};
	\node[left] at ({-1.5*cos(45)},{1.5*sin(45)}) {{\small $t_3$}};
	\node[circle,fill,inner sep=1pt] (d) at ({-1.5*cos(45)},{-1.5*sin(45)}) {};
	\node[left] at ({-1.5*cos(45)},{-1.5*sin(45)}) {{\small $t_4$}};
	\draw[thick] (d) to [bend left=60] node[pos=0.5,below]{{\footnotesize $\Delta_1$}} (a);
	\draw[thick] (b) to [bend left=60] node[pos=0.5,above]{{\footnotesize $\Delta_2$}} (c);
} 
\end{equation}
In this case the energy conservation applies to the $s_0+\omega _1$ energy rather than $s_0$. There are still two independent $\omega _{1,2} $, but this time it is not manifest that the diagram reduces to the product of 2-point functions. Indeed, in general the diagram does not even factorize. However, in the holographic-semiclassical limit, as shown in App.\ \ref{sec:app_uncrossed_4pt_details} we get
\begin{equation}
\langle \cO _1(t_4) \cO _2(t_3) \cO _2(t_2) \cO _1(t_1) \rangle \sim \langle \cO _2(t_3) \cO _2(t_2)\rangle \langle \cO _1(t_4) \cO _1(t_1)\rangle .
\end{equation}
The sign $\sim $ here and below means asymptotic equality.

Therefore these correlation functions are determined by the 2-point functions, just as for a free scalar, and we find that these correlators agree with those of de Sitter.

\section{A brief reminder of OTO correlators in gravity} \label{sec:review_gravity_OTOC}

In this section we review the known behavior of OTO correlators in gravity, in order to be able to compare to it.

\subsection{Anti de Sitter} \label{sec:OTOC_AdS}

We are interested in the out-of-time-order 4-point correlation function which encodes information about the chaotic behavior. The main form of this correlator that is useful to study in a quantum system is by considering two operators, $V$ and $W$, and the correlator
\begin{equation} \label{eq:1_sided_OTO}
\langle VW(t)VW(t)\rangle 
\end{equation}
in a thermal state with inverse temperature $\beta $. We restrict to operators $V$ and $W$ with vanishing expectation values.

In gravity, such correlation functions can be understood as a scattering process \cite{Shenker:2013pqa,Roberts:2014isa,Shenker:2014cwa} and a geometric description in terms of shockwave geometries \cite{Dray:1984ha}.
In holography, it is useful to embed such a system in two copies of the quantum field theory, which we can refer to as `left' and `right', and study it around the thermofield double (TFD) state
\begin{equation}
|\text{TFD}\rangle =\frac{1}{\sqrt{Z}} \sum _n e^{-\beta E_n/2} |n\rangle _L |n\rangle _R,
\end{equation}
where the sum is over a basis of energy eigenstates of the field theory. Each operator $O$ has a corresponding copy in the left system $O_L$ and in the right system $O_R$. With this in mind, we will be interested mainly in two types of correlation functions
\begin{align}
&  \langle V_R W_R(t) V_R W_R(t) \rangle \label{eq:ads_OTO_one_sided} \\
& \langle V_L W_R(t) V_R W_R(t) \rangle \label{eq:ads_OTO_geodesic_config}
\end{align}
in the TFD state.
The configuration \eqref{eq:ads_OTO_one_sided} is a one-sided out-of-time-order 4-point function, and reduces to the OTO 4-point function \eqref{eq:1_sided_OTO} discussed above. This correlator is most easily accessible when considering one copy of the quantum system. Other configurations involving operators on both sides, such as \eqref{eq:ads_OTO_geodesic_config}, sometimes have simper gravitational interpretation as we will recall.
The various OTO correlators usually studied have a simple relation to one another by analytically continuing the time arguments. For instance, \eqref{eq:ads_OTO_geodesic_config} is obtained from \eqref{eq:ads_OTO_one_sided} by adding $-i\beta /2$ to the time argument of the first (leftmost) operator.

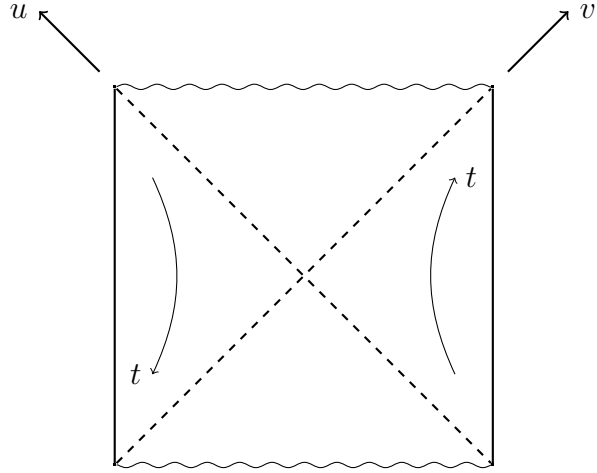
\begin{figure}[h]
\centering
	\begin{tikzpicture}
        \node[fill,inner sep=0pt] (LB) at (0,0) {};
        \node[fill,inner sep=0pt] (LT) at (0,5) {};
        \node[fill,inner sep=0pt] (RB) at (5,0) {};
        \node[fill,inner sep=0pt] (RT) at (5,5) {};
        \draw[thick] (LB)--(LT);
        \draw[thick] (RB)--(RT);
        \draw[thick,dashed] (LB)--(RT);
        \draw[thick,dashed] (LT)--(RB);
        \draw[decorate,decoration={coil,aspect=0,amplitude=1pt, segment length=12.2pt}] (LT)--(RT);
        \draw[decorate,decoration={coil,aspect=0,amplitude=1pt, segment length=12.2pt}] (RB)--(LB);
        \draw[thick,->] (RT) +(0.2,0.2)--+(1,1) node[right] {$v$};
        \draw[thick,->] (LT) +(-0.2,0.2)--+(-1,1) node[left] {$u$};
        \draw[->] (RB) +(-0.5,1.2) to [bend left=25] +(-0.5,3.8) node[right] {$t$};
        \draw[->] (LB) +(0.5,3.8) to [bend left=25] +(0.5,1.2) node[left] {$t$};
        \end{tikzpicture}
\caption{Penrose diagram of a two-sided AdS black hole.}
\label{fig:3d_BTZ}
\end{figure}

In holography we can describe the corresponding bulk process when the dual gravitational theory is a two-sided black hole. One simple case to understand \cite{Shenker:2013pqa} is the configuration \eqref{eq:ads_OTO_geodesic_config} in cases when the scalar field dual to $V$ has large mass compared to the AdS scale. Indeed, exchanging to the symmetric configuration $L \leftrightarrow R$ for convenience, in such a case this correlation function can be viewed as the 2-point function $\langle V_LV_R\rangle _W$ in the state $W_L(t)|\text{TFD}\rangle $ when $W$ is Hermitian (since left and right operators commute).
Before acting with $W_L(t)$ the corresponding bulk geometry in three dimensions is a BTZ metric which is given in Kruskal coordinates by
\begin{equation}
\dd s^2=\frac{-4 \rAdS^2 \dd u \dd v+R^2(1-uv)^2\dd \phi ^2}{(1+uv)^2}
\end{equation}
where $R$ is the horizon radius growing with the mass of the black hole, $R^2=8G_NM\rAdS^2=\left( \frac{2\pi \rAdS^2}{\beta } \right) ^2$, and $\phi$ is an angle (with $2\pi $ periodicity). This describes a black hole with two asymptotic AdS regions, see Fig.\ \ref{fig:3d_BTZ}.

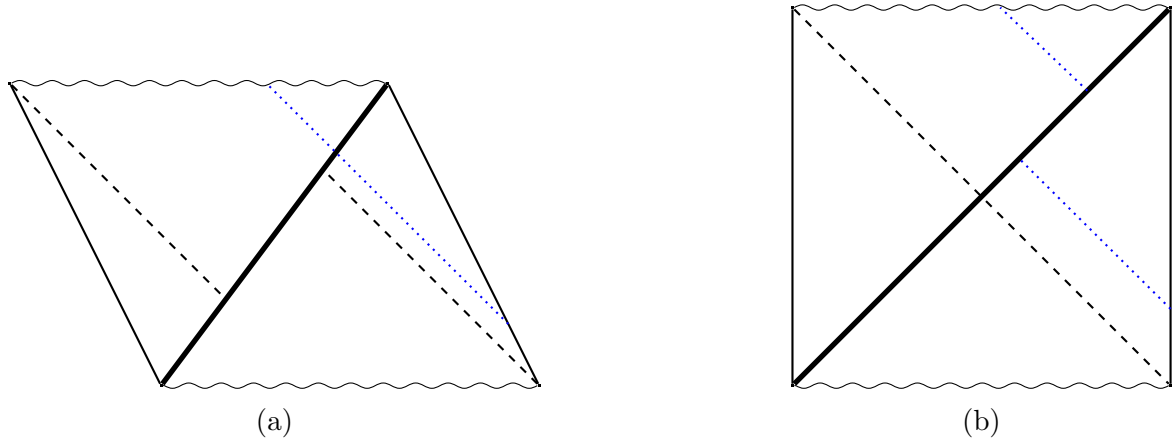
\begin{figure}[ht]
    \centering
    \begin{subfigure}[b]{0.45\textwidth}
        \centering
		\begin{tikzpicture}
                \node[fill,inner sep=0pt] (LB) at (0,0) {};
                \node[fill,inner sep=0pt] (LT) at (-2,4) {};
                \node[fill,inner sep=0pt] (RB) at (5,0) {};
                \node[fill,inner sep=0pt] (RT) at (5-2,4) {};
                \draw[thick] (LB)--(LT);
                \draw[thick] (RB)--(RT) coordinate[pos=0.2] (geodesic_int);
                \draw[line width=2pt] (LB)--(RT);
                \draw[thick,dashed] (RB)--+(-2.8,2.8);
                \draw[thick,dashed] (LT)--+(2.8,-2.8);
                \draw[decorate,decoration={coil,aspect=0,amplitude=1pt, segment length=12.2pt}] (LT)--(RT) coordinate[pos=0.68] (geodesic_end);
                \draw[decorate,decoration={coil,aspect=0,amplitude=1pt, segment length=12.2pt}] (RB)--(LB);
                \draw[thick,dotted, color=blue] (geodesic_int)--(geodesic_end);
                \end{tikzpicture}
        \caption{}
        \label{fig:BTZ_shockwave1}
    \end{subfigure}
    \hfill
    \begin{subfigure}[b]{0.45\textwidth}
        \centering
		\begin{tikzpicture}
                \node[fill,inner sep=0pt] (LB) at (0,0) {};
                \node[fill,inner sep=0pt] (LT) at (0,5) {};
                \node[fill,inner sep=0pt] (RB) at (5,0) {};
                \node[fill,inner sep=0pt] (RT) at (5,5) {};
                \draw[thick] (LB)--(LT);
                \draw[thick] (RB)--(RT) coordinate[pos=0.2] (geodesic_int);
                \draw[line width=2pt] (LB)--(RT);
                \draw[thick,dashed] (LT)--(RB);
                \draw[decorate,decoration={coil,aspect=0,amplitude=1pt, segment length=12.2pt}] (LT)--(RT) coordinate[pos=0.55] (geodesic_end);
                \draw[decorate,decoration={coil,aspect=0,amplitude=1pt, segment length=12.2pt}] (RB)--(LB);
                \draw[thick,dotted, color=blue] (geodesic_int)--(3,3);
                 \draw[thick,dotted, color=blue] (3.9,3.9)--(geodesic_end);
                \end{tikzpicture}
        \caption{}
        \label{fig:BTZ_shockwave2}
    \end{subfigure}
    \caption{Shockwave geometry in AdS and a geodesic propagating on it shown in dotted blue.}
    \label{fig:BTZ_shockwave}
\end{figure}

In the state $W_L(t)|\text{TFD}\rangle $ we add a small amount of energy $E$ on the left side in the past (since $t$ runs backwards on the left hand side). The resulting geometry is simple in the limit that $t$ is large and the energy is small $E \ll M$. Indeed, in such a case, the released particle when non-accelerating would travel close to the past horizon, as it would enter early the black hole. In addition, in the limit, it would not change the mass of the black hole, so the geometry is unchanged on both sides of its trajectory. The resulting geometry is a shockwave geometry. We can write it either in terms of coordinates continuous along the shockwave or coordinates which are not continuous. In terms of the latter, the geometry is
\begin{equation}
\dd s^2= \frac{-4\rAdS^2 \dd u \dd v +4 \rAdS^2 \alpha \delta (u)\dd u^2+R^2(1-uv)^2\dd \phi ^2}{(1+uv)^2} .
\end{equation}
Indeed, it satisfies Einstein's equations with negative cosmological constant and energy-momentum tensor
\begin{equation}
T_{uu} =\frac{\alpha }{4\pi G_N \rAdS^2} \delta (u)
\end{equation}
so that the matter is spherically symmetric with $\alpha  \propto \frac{E}{M} e^{2\pi t/\beta } $ is kept fixed in the limit. We can represent this geometry in two ways differing by how geodesics behave, shown in Fig.\ \ref{fig:BTZ_shockwave}, where Fig.\ \ref{fig:BTZ_shockwave1} in the limit is the continuous geometry. In either case we see that even though one could not send signals between the two sides already before the perturbation, the effect of the shockwave is such that it becomes only harder to do so.

\begin{figure}[ht]
    \centering
    \begin{subfigure}[b]{0.45\textwidth}
        \centering
		\begin{tikzpicture}
                \node[fill,inner sep=0pt] (LB) at (0,0) {};
                \node[fill,inner sep=0pt] (LT) at (0,5) {};
                \node[fill,inner sep=0pt] (RB) at (5,0) {};
                \node[fill,inner sep=0pt] (RT) at (5,5) {};
                \draw[thick] (LB)--(LT) coordinate[pos=0.5] (left_V);
                \draw[thick] (RB)--(RT) coordinate[pos=0.5] (right_V);
                \draw[line width=2pt] (LB)--(RT) coordinate[pos=0.6] (geodesic_point1) coordinate[pos=0.4] (geodesic_point2);
                \draw[thick,dashed] (LT)--(RB);
                \draw[decorate,decoration={coil,aspect=0,amplitude=1pt, segment length=12.2pt}] (LT)--(RT) coordinate[pos=0.55] (geodesic_end);
                \draw[decorate,decoration={coil,aspect=0,amplitude=1pt, segment length=12.2pt}] (RB)--(LB);
                \draw[thick,dotted, color=blue] (left_V)--(geodesic_point1);
                 \draw[thick,dotted, color=blue] (geodesic_point2)--(right_V);
                 \node[left] at (left_V) {$V_L$};
                 \node[right] at (right_V) {$V_R$};
                \end{tikzpicture}
        \caption{AdS}
        \label{fig:geodesic_length_AdS}
    \end{subfigure}
    \hfill
    \begin{subfigure}[b]{0.45\textwidth}
        \centering
		\begin{tikzpicture}
                \node[fill,inner sep=0pt] (LB) at (0,0) {};
                \node[fill,inner sep=0pt] (LT) at (0,5) {};
                \node[fill,inner sep=0pt] (RB) at (5,0) {};
                \node[fill,inner sep=0pt] (RT) at (5,5) {};
                \draw[thick] (LB)--(LT) coordinate[pos=0.5] (left_V);
                \draw[thick] (RB)--(RT) coordinate[pos=0.5] (right_V);
                \draw[line width=2pt] (LB)--(RT) coordinate[pos=0.4] (geodesic_point1) coordinate[pos=0.6] (geodesic_point2);
                \draw[thick,dashed] (LT)--(RB);
                \draw[thick] (LT)--(RT) coordinate[pos=0.55] (geodesic_end);
                \draw[thick] (RB)--(LB);
                \draw[thick,dotted, color=blue] (left_V)--(geodesic_point1);
                 \draw[thick,dotted, color=blue] (geodesic_point2)--(right_V);
                 \node[left] at (left_V) {$V_L$};
                 \node[right] at (right_V) {$V_R$};
                \end{tikzpicture}
        \caption{dS}
        \label{fig:geodesic_length_dS}
    \end{subfigure}
    \caption{Geodesic length in the shockwave geometry.}
    \label{fig:geodesic_length}
\end{figure}
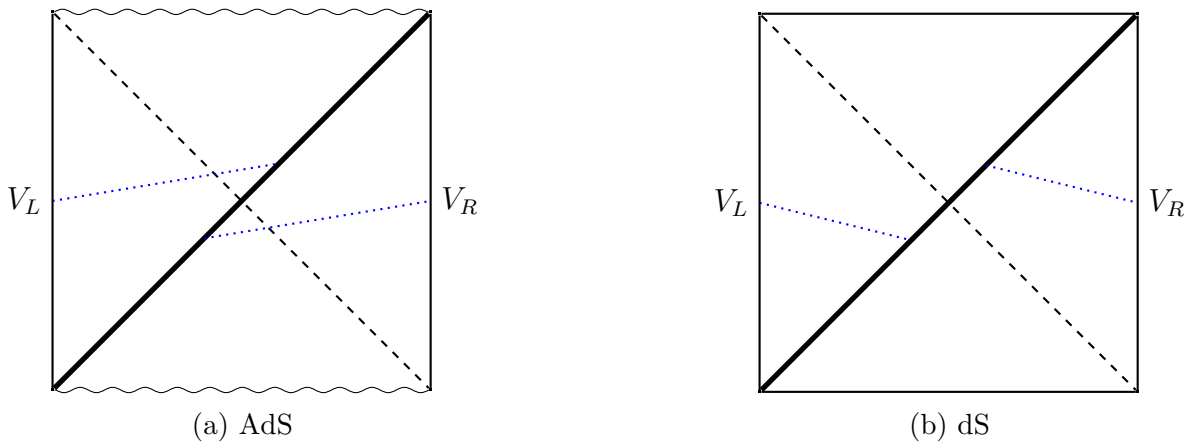

The 2-point function $\langle V_LV_R\rangle _W$ in a free theory is given by a path integral over curves connecting the two points $\int DX \, e^{-im_V L(X)} $ where $L(X)$ is the length of the curve and $m_V$ is the mass of the field dual to $V$. In the limit of large $m_V$, for spacelike separated points, this is well approximated by $e^{-m_V D} $ where $D$ is the (renormalized) geodesic distance between the two points. When drawing the spacetime as in Fig.\ \ref{fig:BTZ_shockwave2}, geodesics oriented as in the positive $u$ direction are shifted by an amount $\Delta v=\alpha $. Therefore, the geodesic distance dominating this 2-point function is as shown in Fig.\ \ref{fig:geodesic_length_AdS}, which clearly becomes bigger compared to the unperturbed geometry and the 2-point function becomes smaller. This intuitively makes sense since the two sides become even less connected.

In this description the correlation function of the $V$ operators is evaluated in a geometry affected by the $W$ operators with a given $\alpha $, which is appropriate when $W$ is much heavier than $V$.
More generally, these OTO 4-point functions are described in the bulk by 2-2 scattering, where $\alpha $ is integrated over. This is clear in the alternative configuration
\begin{equation} \label{eq:ads_OTO_two_sided}
\langle V_L W_R(t) V_R W_L(t)\rangle = \langle V_L(-t/2) W_R(t/2) V_R(-t/2) W_L(t/2)\rangle
\end{equation}
as $V_R(-t/2) W_L(t/2)$ creates an in-state with $W$ on the LHS and $V$ on the RHS in the far past, and the rest gives the corresponding out state. However, other configurations give a similar scattering picture.

In any case, the OTO 4-point function goes to zero exponentially in time at large enough times. The way it goes to zero depends on the details of the problem, such as which operators we use, whether the shockwave is spherical or localized, and so on.  At earlier times which are long compared to the inverse temperature but much shorter than the scrambling time, the various OTO correlators behave as
\begin{equation} \label{eq:OTO_leading_behavior}
\frac{\langle VW(t)VW(t)\rangle }{\langle VV\rangle \langle WW\rangle } \sim 1 - f e^{2\pi t/\beta } +\cdots 
\end{equation}
where we left implicit the precise configuration. Generally, further regularization is needed to separate the times of the two $V$ operators and similarly for the times of the $W$ operators. This can be done by at least slightly separating the time difference between the first and second operator of the same sort by a negative imaginary value. In the configuration \eqref{eq:ads_OTO_one_sided} all operators are on the same side and so we should do this regularization for both pairs. In this case $f$ is a small purely imaginary coefficient proportional to $G_N$.
In the configuration \eqref{eq:ads_OTO_geodesic_config}, as mentioned the first $V$ is shifted along half the thermal circle, so no regularization is needed for it, and we can regularize the $W$ as just discussed, resulting in a real positive $f$. This is as expected from the geodesic distance argument, the correlation decreasing because of the perturbation.
% (In the two-sided case, \eqref{eq:ads_OTO_two_sided}, the coefficient is again purely imaginary.)

\subsection{de Sitter} \label{sec:OTOC_dS}

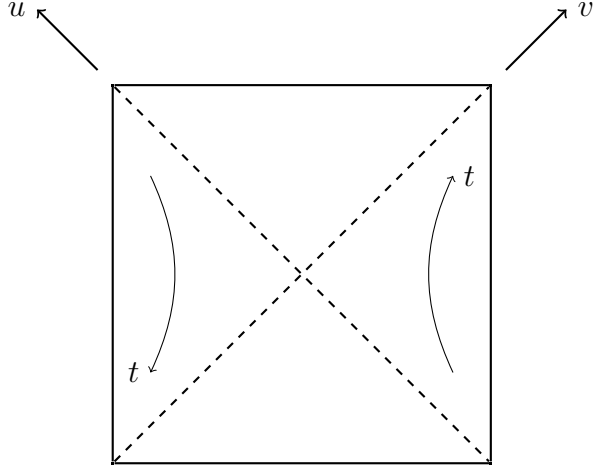
\begin{figure}[h]
\centering
	\begin{tikzpicture}
        \node[fill,inner sep=0pt] (LB) at (0,0) {};
        \node[fill,inner sep=0pt] (LT) at (0,5) {};
        \node[fill,inner sep=0pt] (RB) at (5,0) {};
        \node[fill,inner sep=0pt] (RT) at (5,5) {};
        \draw[thick] (LB)--(LT);
        \draw[thick] (RB)--(RT);
        \draw[thick,dashed] (LB)--(RT);
        \draw[thick,dashed] (LT)--(RB);
        \draw[thick] (LT)--(RT);
        \draw[thick] (RB)--(LB);
        \draw[thick,->] (RT) +(0.2,0.2)--+(1,1) node[right] {$v$};
        \draw[thick,->] (LT) +(-0.2,0.2)--+(-1,1) node[left] {$u$};
        \draw[->] (RB) +(-0.5,1.2) to [bend left=25] +(-0.5,3.8) node[right] {$t$};
        \draw[->] (LB) +(0.5,3.8) to [bend left=25] +(0.5,1.2) node[left] {$t$};
        \end{tikzpicture}
\caption{Penrose diagram of global de Sitter space.}
\label{fig:de_Sitter_Penrose}
\end{figure}

The situation in de Sitter space is analogous, but interestingly there are important differences \cite{Aalsma:2020aib,Anninos:2018svg,Blommaert:2020tht}. The Penrose diagram of de Sitter looks similar and is shown in Fig.\ \ref{fig:de_Sitter_Penrose}. One of the main difficulties of dS is the lack of a timelike non-gravitating asymptotic boundary that can be helpful in understanding quantum gravity around such a space. Indeed, global dS has spherical spatial slices. In the Penrose diagram, the right and left wedges are known as `static patches' having a timelike Killing vector. The right and left edges of the diagram are the poles of the sphere where it is common to position static observers. Similarly to a black hole, a static patch has a horizon, which is a cosmological horizon originating in the expansion of the space. Signals falling into the horizon have blueshifted energies compared to the static observer, just as objects falling into a black hole have such a blueshift compared to an asymptotic observer.

We can similarly cover the entire space with Kruskal coordinates, such that in three dimensions the metric is given by
\begin{equation}
\dd s^2 = \rdS^2 \frac{-4 \dd u \dd v+(1+uv)^2\dd \phi ^2}{(1-uv)^2} 
\end{equation}
where $\rdS$ is the dS radius and the inverse temperature associated to the horizon is $\beta =2\pi \rdS$. In the static patches one could use static coordinates $t$, $r$, $\phi $ analogous to the Schwarzschild coordinates. By the relation of these coordinates to global coordinates one could see that taking $t \to t-i\pi \rdS$ takes a point to its antipodal point (with respect to the center of the Penrose diagram, leaving the internal angular coordinate invariant).

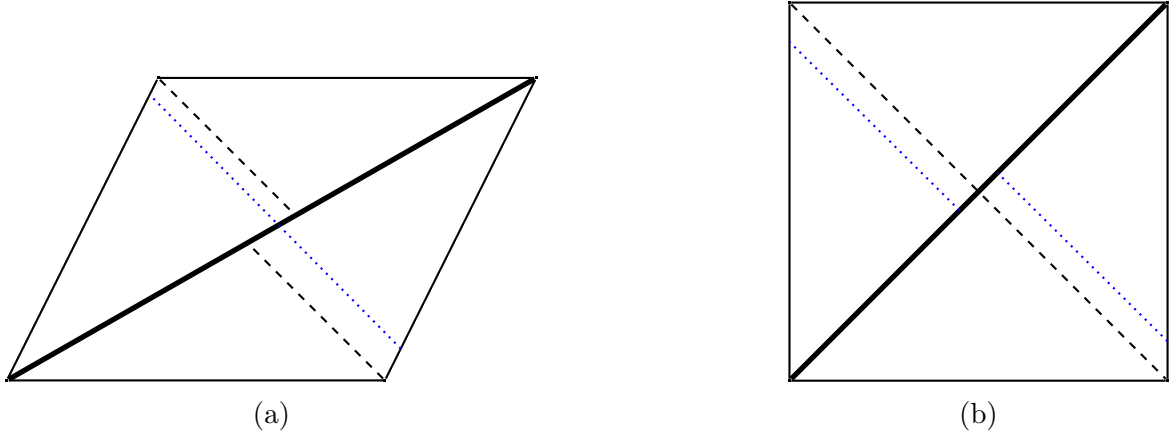
\begin{figure}[ht]
    \centering
    \begin{subfigure}[b]{0.45\textwidth}
        \centering
		\begin{tikzpicture}
                \node[fill,inner sep=0pt] (LB) at (0,0) {};
                \node[fill,inner sep=0pt] (LT) at (2,4) {};
                \node[fill,inner sep=0pt] (RB) at (5,0) {};
                \node[fill,inner sep=0pt] (RT) at (5+2,4) {};
                \draw[thick] (LB)--(LT) coordinate[pos=0.95] (geodesic_end);
                \draw[thick] (RB)--(RT) coordinate[pos=0.1] (geodesic_int);
                \draw[line width=2pt] (LB)--(RT);
                \draw[thick,dashed] (RB)--+(-1.8,1.8);
                \draw[thick,dashed] (LT)--+(1.8,-1.8);
                \draw[thick] (LT)--(RT);
                \draw[thick] (RB)--(LB);
                \draw[thick,dotted, color=blue] (geodesic_int)--(geodesic_end);
                \end{tikzpicture}
        \caption{}
        \label{fig:dS_shockwave1}
    \end{subfigure}
    \hfill
    \begin{subfigure}[b]{0.45\textwidth}
        \centering
		\begin{tikzpicture}
                \node[fill,inner sep=0pt] (LB) at (0,0) {};
                \node[fill,inner sep=0pt] (LT) at (0,5) {};
                \node[fill,inner sep=0pt] (RB) at (5,0) {};
                \node[fill,inner sep=0pt] (RT) at (5,5) {};
                \draw[thick] (LB)--(LT) coordinate[pos=0.9] (geodesic_end);
                \draw[thick] (RB)--(RT) coordinate[pos=0.1] (geodesic_int);
                \draw[line width=2pt] (LB)--(RT);
                \draw[thick,dashed] (LT)--(RB);
                \draw[thick] (LT)--(RT);
                \draw[thick] (RB)--(LB);
                \draw[thick,dotted, color=blue] (geodesic_int)--(2.75,2.75);
                 \draw[thick,dotted, color=blue] (2.25,2.25)--(geodesic_end);
                \end{tikzpicture}
        \caption{}
        \label{fig:dS_shockwave2}
    \end{subfigure}
    \caption{Shockwave geometry in dS and a geodesic propagating on it shown in dotted blue.}
    \label{fig:dS_shockwave}
\end{figure}

Similarly to AdS, there is a shockwave solution (see also, e.g., \cite{Hotta:1992qy,Hotta:1992wb,Sfetsos:1994xa}) corresponding to a particle released far in the past
\begin{equation}
\dd s^2 = \rdS^2 \bigg[
-\frac{4}{(1-uv)^2} \dd u\dd v-4\alpha \delta (u)\dd u^2+\frac{(1+uv)^2}{(1-uv)^2} \dd \phi ^2 \bigg]
\end{equation}
where the EM tensor is
\begin{equation}
T_{uu} =\frac{\alpha }{4\pi G_N \rdS^2} \delta (u)
\end{equation}
and so again $\alpha >0$ corresponds to positive energies. An important difference from AdS is that geodesics oriented as in the positive $u$ direction are shifted by an amount $\Delta v=-\alpha $, that is, time advance, the opposite sign compared to AdS. This is important since while before the perturbation it is impossible to send signals between two observers sitting at the antipodal poles, after the shockwave one could send signals to the other if done early enough.
Recalling that the reason there is a horizon to begin with is an accelerated expansion of the universe making it impossible to send signals everywhere, and is caused by the negative pressure of the cosmological constant, sending matter that has ordinary matter properties slows this expansion and makes it possible to send signals further.
The shockwave geometry is shown in Fig.\ \ref{fig:dS_shockwave}.

The analogue of left and right operators here are operators located at the left and right poles. Applying the conclusion on time advance to the correlator \eqref{eq:ads_OTO_geodesic_config} in the geodesic approximation, the corresponding geodesic in the perturbed geometry is shown in Fig.\ \ref{fig:geodesic_length_dS} which clearly shows that it becomes shorter compared to the unperturbed geometry, and so the correlator should \emph{grow} the earlier we apply the perturbation. This makes sense since it is easier to send a signal between the two sides. So in this configuration the correlator is approximately real and \eqref{eq:OTO_leading_behavior} should have \emph{negative} $f$.

For the single sided correlator \eqref{eq:ads_OTO_one_sided} two differences have been found compared to AdS \cite{Aalsma:2020aib}. First, the leading time dependence is $e^{4\pi t/\beta } $ with a \emph{doubled} exponent. Second, the coefficient $f$ is real and positive rather than purely imaginary.

Note that a geodesic approximation calculation of \eqref{eq:ads_OTO_geodesic_config} (as in \cite{Aalsma:2020aib}) gives an oscillatory behavior of the correlator for very long times. Such an approximation is only valid for masses that are in particular in the principal series representation.

\section{Crossed 4-point function} \label{sec:OTOC_4pf}

Up to now in the microscopic model we discussed the classical limit of gravity. Indeed, Newton's constant has made no explicit appearance. Our goal is to have our principle for describing a gravitational system agree with all the known results on the semiclassical limit of gravity. One could wonder how to approach small corrections in Newton's constant. In the framework we discuss of describing gravity using a heavy object as a probe, a natural object to achieve this is by using out-of-time-order correlation functions. % \footnote{Note in particular that there is no obvious reason why the entropy of the quantum mechanical system would agree with the expected entropy of the static patch. If such a relation exists, it would be interesting to understand it.} 
We saw in Sec.\ \ref{sec:review_gravity_OTOC} that when considering times of the order of the scrambling time in gravity, we are probing small corrections in Newton's constant. In this section we discuss the OTO 4-point function. It is most clean to consider two species of operators $\cO _1$ and $\cO _2$ and the 4-point function
\begin{equation}
\langle \cO _2(t_4) \cO _1(t_3) \cO _2(t_2) \cO _1(t_1) \rangle .
\end{equation}
In terms of diagrams, there is a single diagram contributing to this correlation function which is a crossed diagram
\begin{equation} \label{eq:4pf_crossed}
\tikz[baseline=0ex]{
	\draw[thick] (1.5,0) arc (0:360:1.5)
		node[pos=0,right]{{\small $s_0+\omega_1 $}}
		node[pos=0.25,above]{{\small $s_0 +\omega_2$}}
		node[pos=0.5,left]{{\small $s_0+\omega_3 $}}
		node[pos=0.75,below]{{\small $s_0 $}};
	\node[circle,fill,inner sep=1pt] (a) at ({1.5*cos(45)},{-1.5*sin(45)}) {};
	\node[right] at ({1.5*cos(45)},{-1.5*sin(45)}) {{\small $\, t_1$}};
	\node[circle,fill,inner sep=1pt] (b) at ({1.5*cos(45)},{1.5*sin(45)}) {};
	\node[right] at ({1.5*cos(45)},{1.5*sin(45)}) {{\small $\, t_2$}};
	\node[circle,fill,inner sep=1pt] (c) at ({-1.5*cos(45)},{1.5*sin(45)}) {};
	\node[left] at ({-1.5*cos(45)},{1.5*sin(45)}) {{\small $t_3$}};
	\node[circle,fill,inner sep=1pt] (d) at ({-1.5*cos(45)},{-1.5*sin(45)}) {};
	\node[left] at ({-1.5*cos(45)},{-1.5*sin(45)}) {{\small $t_4$}};
	\draw[thick] (a)--(c) node[pos=0.75,left]{{\footnotesize $\Delta_1$}};
	\draw[thick] (b)--(d) node[pos=0.25,right]{{\footnotesize $\, \Delta_2$}};
} 
\end{equation}
The diagrammatic rules discussed before are actually not sufficient in order to write the expression for this diagram. In addition to the rules already mentioned, there is one additional rule which requires us to assign a value to the intersection of chords in the bulk of the diagram \cite{Berkooz:2018jqr}. This value is in fact the $6j$ symbol of a $q$-deformation of $\mathfrak{sl}_2$. In this section we discuss this correlation function while leaving the details of the calculation to the appendices.

\subsection{One copy}

For orientation, let us start with a single copy of our microscopic system. In the semiclassical limit we consider, this system is holographic and describes gravity in AdS space. Some details of the analysis are discussed in App.\ \ref{sec:app_one_copy_crossed_4pf}. The operators are just the operators \eqref{eq:single_side_operators} in a single copy labeled by a single parameter $\Delta $. The state is of temperature that we denote by $T_1$ to distinguish it from the temperature of the two copy system.

As mentioned, we are interested in times of order the scrambling time. More precisely, we take $t_2,t_4 \sim t$ and $t_1,t_3 \sim 0$ up to finite times, while $t$ and $s_0$ are taken to be large while keeping $\frac{1}{s_0} e^{2\pi T_1t} $ finite. In this case we can write the full result in the semiclassical limit and it is given by \eqref{eq:one_copy_OTO_result} which we reproduce here
\begin{equation} \label{eq:one_copy_OTO_4pf}
\begin{split}
& \frac{\langle \cO _2(t_4) \cO _1(t_3) \cO _2(t_2) \cO _1(t_1) \rangle \onecopy}{\langle \cO _2(t_4) \cO _2(t_2)\rangle \onecopy \langle \cO _1(t_3) \cO _1(t_1)\rangle \onecopy } = x^{2\Delta _1} U(2\Delta _1,1+2\Delta _1-2\Delta _2,x) \\
& \qquad \text{with} \quad x=8s_0i \sinh(\pi T_1t_{31} )\sinh(\pi T_1t_{42} )e^{\pi T_1 (t_1+t_3-t_2-t_4)} 
\end{split}
\end{equation}
where we wrote the result for the 4-point function normalized by a product of 2-point functions, and the confluent hypergeometric function $U$ is defined in \eqref{eq:U_func_def}. We have also added a superscript to indicate that these are single-copy correlators. This result has appeared several times previously in the literature, such as in \cite{Maldacena:2016upp,Lam:2018pvp,Mukhametzhanov:2023tcg}.

The general structure of an OTO 4-point function is as follows. At very long times, all operators are very much separated from each other, and we expect the correlator to go to zero. This is indeed the case for \eqref{eq:one_copy_OTO_4pf} as can be seen by considering the $x \to 0$ limit, as done in App.\ \ref{sec:app_one_copy_crossed_4pf}.

In contrast, times in the regime we discussed, but still much shorter than the scrambling time, are the range describing the Lyapunov chaotic behavior. In this range the OTO correlator can be described using an expansion in $\frac{1}{s_0} e^{2\pi T_1t} $. This description is simply the large $x$ expansion of \eqref{eq:one_copy_OTO_4pf}. As shown in \eqref{eq:app_one_copy_OTO_expansion}, this expansion up to the first subleading order is
\begin{equation} \label{eq:one_copy_OTO_expansion}
\frac{\langle \cO _2(t_4) \cO _1(t_3) \cO _2(t_2) \cO _1(t_1) \rangle \onecopy}{\langle \cO _2(t_4) \cO _2(t_2)\rangle \onecopy \langle \cO _1(t_3) \cO _1(t_1)\rangle \onecopy } \sim 1+ \frac{i}{2} \frac{\Delta _1\Delta _2}{s_0 \sinh(\pi T_1 t_{31} ) \sinh(\pi T_1 t_{42} )} e^{\pi T_1(t_2+t_4-t_1-t_3)} + \cdots
\end{equation}
(even though one could write the expansion to any order).
In writing \eqref{eq:one_copy_OTO_expansion} we have kept the full dependence on the four times. The reason for this is that generally taking $t_2=t_4=t$ and $t_1=t_3=0$ is really a singular limit and one needs to regularize it by slightly separating the times.

Let us compare this to the expectation in AdS, as reviewed in Sec.\ \ref{sec:OTOC_AdS}. 
To achieve the configuration \eqref{eq:ads_OTO_one_sided} which is one sided, we should in principle just take $t_2=t_4=t$ and $t_1=t_3=0$, but as we see in \eqref{eq:one_copy_OTO_expansion} we need further regularization so we take $t_1=0$, $t_2=t$, $t_3=-i\epsilon $, and $t_4=t-i\epsilon $ for a small regularizing $\epsilon $. In this case \eqref{eq:one_copy_OTO_expansion} indeed takes the form \eqref{eq:OTO_leading_behavior} with a purely imaginary $f$, just as in AdS. If instead we are interested in the configuration \eqref{eq:ads_OTO_geodesic_config} we can move the rightmost operator halfway through the thermal circle, that is $t_1=i/(2T_1)$, $t_2=t$, $t_3=0$, and $t_4=t-i\epsilon $. Note that because of the imaginary part of $t_1$ we do not need a further regularization there. From the holographic point of view this gives essentially $\langle \cO _{1,L} \cO _{2,R} (t-i\epsilon ) \cO _{1,R} \cO _{2,R} (t)\rangle $, which is precisely a regularized \eqref{eq:ads_OTO_geodesic_config}. In \eqref{eq:one_copy_OTO_expansion} we see that this results in a real and positive $f$. This is the expected result in AdS as discussed in Sec.\ \ref{sec:OTOC_AdS} caused by the time delay in the presence of a shockwave.

\subsection{Two copies}

We are again interested in the same OTO correlation function associated with temperature $T$, and the same behavior of the four time variables.

We can do a similar analysis in the two-copy system. In principle, one could try to evaluate the crossed 4-point function in full form (in the semiclassical limit) and get an expression analogous to \eqref{eq:one_copy_OTO_4pf}. We have not done this, and it would be interesting to get such a nice description which we leave to future work. Instead, we have a more complicated integral expression written explicitly in \eqref{eq:crossed_4pf_int_expr_before_simplify}. We can still analyze this expression and understand the physics that it encodes.

The first question one could wonder about is what is the behavior of the OTO 4-point function for very long times, far after the scrambling time $t \sim T^{-1} \log s_0$ (but much before the recurrence time). This is non-trivial because the operators we consider \eqref{eq:doubled_3d_ops} are integrated operators, non-local in time. It is therefore not clear whether such operators would give OTO correlators that go to zero after a long time. In fact, we have checked this explicitly starting from the integral expression and have shown in App.\ \ref{sec:app_crossed_4pf_details} that even though the operators are integrated operators, the OTOC still decays to zero after very long time. In this particular sense the operators behave nicely, even though it is not obvious from their definition.

To compare this situation to de Sitter space, we note that operators in our microscopic system should have real and positive $\Delta $'s. Translating this to scalars in the bulk of de Sitter space, this means we are in principle considering masses in the complementary series representation. In this case the OTOC does decay for long times.

Next we would like to understand the chaotic behavior which is encoded at early scrambling times. We can systematically develop an expansion in exponentials of $t$, analogous to \eqref{eq:one_copy_OTO_expansion}. We do not start from a rather simple function and just expand it as in the single copy system, but rather do it from the integral expression. Technical details are given in App.\ \ref{sec:app_crossed_4pf_details}. We find that
\begin{equation} \label{eq:crossed_4pf_exp}
\begin{split}
& \frac{\langle \cO _2(t_4) \cO _1(t_3) \cO _2(t_2) \cO _1(t_1) \rangle}{\langle \cO _2(t_4) \cO _2(t_2)\rangle \langle \cO _1(t_3) \cO _1(t_1)\rangle } = 
\frac{\log s_0}{4\pi ^2T} \bigg\{ 1
-\frac{e^{4\pi T t_{43} } }{4s_0^2} \ratiocoeff_{\Delta _1}  (2\pi T t_{31} ) \cdot \ratiocoeff '_{\Delta _2}  (2\pi Tt_{42} ) + \cdots \bigg\}
\end{split}
\end{equation}
where
\begin{equation} \label{eq:g_def}
\ratiocoeff_{\Delta } (t) =\frac{1}{2} +\Big( \Delta -\frac{1}{2} \Big) \coth \left( \mu (t+i\pi )\right) +\frac{1}{2} \coth t+\Big( \Delta -\frac{1}{2} \Big) \coth \left( \mu (t+i\pi )\right)  \coth (t) +\frac{1}{2} \left( \sinh(t)\right) ^{-2}
\end{equation}
and
\begin{equation} \label{eq:g_prime_def}
\ratiocoeff '_{\Delta } (t) =\frac{1}{2} -\Big( \Delta -\frac{1}{2} \Big) \coth \left( \mu (t+i\pi )\right) -\frac{1}{2} \coth t+\Big( \Delta -\frac{1}{2} \Big) \coth \left( \mu (t+i\pi )\right)  \coth (t) +\frac{1}{2} \left( \sinh(t)\right) ^{-2}
\end{equation}
and again in both cases $\mu $ is directly related to $\Delta $ via $\mu =1-2\Delta $.

We now compare this to gravitational physics. The times corresponding to the different configurations are the same as discussed in the case of a single copy, and we use the same regularization. 

We emphasize that in our setup of a worldline object used to study gravity, we should really consider only the one-sided configuration \eqref{eq:ads_OTO_one_sided}. Indeed we have only one such object that in principle knows nothing about ``the other side''.
Let us examine then \eqref{eq:crossed_4pf_exp} for the time configuration $t_1=0$, $t_2=t$, $t_3=-i\epsilon $, and $t_4=t-i\epsilon $. First, we see that the Lyapunov exponent is doubled, that is, the subleading term goes as $e^{4\pi Tt} $ rather than $e^{2\pi Tt} $. Second, with this regularization, the dominant term in $\ratiocoeff,\ratiocoeff '$ is the last term in \eqref{eq:g_def} and \eqref{eq:g_prime_def} giving $-\frac{1}{2(2\pi T \epsilon) ^2} $ meaning that the relative coefficient between the leading and subleading terms is real and negative. So $f$ in \eqref{eq:OTO_leading_behavior} is real and positive this time. All of these features, the doubled exponent and $f$ being real and positive, match exactly with the known behavior of de Sitter, as reviewed in Sec.\ \ref{sec:OTOC_dS}.

The rationale behind the analytic continuation of time to go from one pole to the other in de Sitter can be explained by the relation between the static time coordinate and coordinates covering the entire space. Clearly in our construction time is only defined along a single worldline and has no such obvious relation. This is another reason to consider only one-sided configurations. For completeness, however, let us still consider the analogue of the configuration \eqref{eq:ads_OTO_geodesic_config} with times $t_1=i/(2T)$, $t_2=t$, $t_3=0$, and $t_4=t-i\epsilon $. In fact, it seems that in this configuration there is still a divergence in $\ratiocoeff_{\Delta _1} (2\pi Tt_{31} )\to \ratiocoeff_{\Delta _1} (-i\pi )$ even though the two times are separated. This is an artifact of the way we wrote the expression. Indeed, regularizing this slightly to $\ratiocoeff_{\Delta _1} (-i\pi -i\epsilon )$ gives $\frac{2}{3} \Delta_1 (1-\Delta_1)$ to leading order, independent of $\epsilon $. The function $\ratiocoeff '_{\Delta _2} (2\pi Tt_{42} )$ is regularized and again is real and negative. So in total the relative coefficient between the leading and subleading terms in \eqref{eq:crossed_4pf_exp} is positive. This means that $f$ in the analogue of \eqref{eq:OTO_leading_behavior} is \emph{negative}. This is the de Sitter phenomenon of time advance and being able to send signals between the two sides.\footnote{Note that restricting to the real part of the OTOC in dS, while the geodesic approximation corresponding to the principal series gives a non-doubled leading exponent, the case of conformally coupled scalars which are in the complementary series analyzed in \cite{Aalsma:2020aib} seems to give a doubled leading exponent.}

Let us comment on a further non-trivial point. We have seen that in the semiclassical limit, both uncrossed 4-point functions \eqref{eq:4pf_uncrossed_diagram_1} and \eqref{eq:4pf_uncrossed_diagram_2} factorize into a product of 2-point functions. For any large enough finite time (independent of the semiclassical parameter), we would like to have the same property for the crossed channel \eqref{eq:4pf_crossed}. This is indeed the case for all 4-point functions in the single copy system. For two copies, this is analyzed in App.\ \ref{sec:finite_times}. We find interestingly that the functional dependence on the times in the semiclassical limit is the same as that of the product of 2-point functions  $\langle \cO _2(t_4) \cO _2(t_2)\rangle \langle \cO _1(t_3) \cO _1(t_1)\rangle$. This happens in a rather non-trivial way, the details of which are shown in App.\ \ref{sec:finite_times}. However, strictly speaking there is a coefficient in front of the product of 2-point functions which is $\frac{\log s_0}{4\pi ^2T} $. This agrees with the leading short time behavior in the scrambling regime as can be seen in \eqref{eq:crossed_4pf_exp}. This coefficient is naturally thought of as being small since $T \sim s_0$ is large. It is interesting to understand the overall coefficient in the crossed channel. In particular, there is in fact an additional factor of $q^{\Delta _1\Delta _2} $ \cite{Berkooz:2018jqr} that we have not written, and does not contribute in the semiclassical limit. The functional form thus obeys large $N$ factorization, but not necessarily the coefficient. One direction to understanding this is to notice that the crossed channel has three independent momenta $\omega _i$ in contrast to the uncrossed channels. This is important when considering the projection of states in the two copies. Choosing a different $\projJ$ would not affect the ratio of the uncrossed 4-point functions to the product of 2-point functions, but would affect the same ratio for the crossed 4-point function. In particular, it will change the constant in the factorization of the crossed correlator. One might in principle choose the constant to give a unit coefficient in this relation. Note that this would mean that we are introducing a dependence on a coupling of the system, which can be phrased as $\JMS$ in units of the temperature, in the definition of the system.

\subsubsection{Perturbative expansion} \label{sec:4pf_expansion}

In this subsection we would like to write the extension of \eqref{eq:crossed_4pf_exp} to the full perturbative expansion. The result might be slightly awkward when written for the particular operators \eqref{eq:doubled_3d_ops}, so we write it for the more general operators \eqref{eq:doubled_generic_ops}. These have a simple interpretation. Expressions such as \eqref{eq:micro_2pf_3d} for the 2-point function are reminiscent of the 2-point function in de Sitter of general dimension $d$, and so we can use such operators to describe more general bulk dimensions.\footnote{This was first noticed by Alexey Milekhin \cite{MilekhinXuInProgress}.} Indeed, parametrizing the two parameters appearing in \eqref{eq:doubled_generic_ops} by
\begin{equation} \label{eq:dimensions_giving_dS_d}
2\Delta ^{(1),(2)} =\frac{d-1}{2} \pm i\mud
\end{equation}
the 2-point function of such an operator is proportional to the 2-point function of a scalar in $d$-dimensional de Sitter space $G_{d,\mud}$ (written explicitly in \eqref{eq:2pf_dS_2F1}):
\begin{equation}
\begin{split}
& \langle \cO (t) \cO (0)\rangle = \frac{(2\lambda s_0)^{d-1}  \Gamma \left( \frac{d}{2} \right) \Gamma \left( \frac{d-1}{2} \right) ^2}{2\pi ^{d/2} T^{d-1} \, \Gamma (d-1)\Gamma (2\Delta ^{(1)} )\Gamma (2\Delta ^{(2)} )}  G_{d,\mud } (t)
\end{split}
\end{equation}
where $\mud $ is related to the mass $m$ of the scalar by $\mud =\sqrt{m^2\rdS^2-\left( \frac{d-1}{2} \right) ^2}$. (For $d=3$ our notation $\mu ^2$ corresponds to $(-\mud^2)$; in order to get real $\Delta ^{(1),(2)} $ we need an imaginary $\mud$.)
A derivation of this is given in App.\ \ref{sec:app_2pt_relation_dS}.

Consider now two operators $\cO _1$ and $\cO _2$ each of the form \eqref{eq:doubled_generic_ops}, choosing their parameters to correspond to mass parameters $\mud_1$ and $\mud_2$ and to $d$-dimensional de Sitter space. As we show in App.\ \ref{sec:crossed_pert_evaluation_general_d}, we can express the perturbative expansion of the 4-point function in terms of the 2-point functions $G_{d,\mud} (t)$ in de Sitter space. To this end, we need to use such 2-point functions with various dimensions, but fixed $\mud_1$ and $\mud_2$ (which would correspond to varying masses in the different dimensions). We find the following expansion for the normalized 4-point function
\begin{equation}
\begin{split}
& \frac{\langle \cO _2(t_4) \cO _1(t_3) \cO _2(t_2) \cO _1(t_1) \rangle}{\langle \cO _2(t_4) \cO _2(t_2)\rangle \langle \cO _1(t_3) \cO _1(t_1)\rangle }= \frac{\log s_0}{4\pi ^2 \, T} 
\sum _{n=0} ^{\infty } \frac{(-1)^n}{(n!)^2} \frac{1}{(2s_0)^{2n}} e^{2\pi nT(2t_{43} +t_{31} -t_{42} )} \\
& \qquad \qquad \frac{\Gamma \left( \frac{d+2n-1}{2} \right) ^4 \Gamma \left( \frac{d}{2} +n\right) ^2}{\pi ^{2n} T^{4n}\, \Gamma (d+2n-1)^2} \frac{\Gamma (d-1)^2}{\Gamma \left( \frac{d-1}{2} \right) ^4 \Gamma \left( \frac{d}{2} \right) ^2}  \frac{G_{d+2n,\mud _1} (t_{31} )G_{d+2n,\mud _2} (t_{42} )}{G_{d,\mud _1} (t_{31} )G_{d,\mud _2} (t_{42} )} .
\end{split}
\end{equation}
Eq.\ \eqref{eq:crossed_4pf_exp} captures the $n=0,1$ terms of this expression for $d=3$.

\subsection{Comments on the bound on chaos}

The non-trivial behavior of the OTOC raises questions about tension with the bound on chaos. We address this here, starting first with a few comments.

In characterizing chaos one could use two related quantities, which are the squared commutator
\begin{equation}
C(t) = -\langle [W(t),V]^2\rangle 
\end{equation}
and the OTO 4-point function
\begin{equation}
F(t) = \langle VW(t)VW(t)\rangle 
\end{equation}
where we assume that $V$ and $W$ are Hermitian operators with vanishing expectation value, and all expectation values are in the thermal state. These two indeed are related by
\begin{equation}
C(t) = \langle W(t)VVW(t)\rangle + \langle VW(t)W(t)V\rangle - 2 \Re F(t)
\end{equation}
when $t$ is real.

A relevant time scale is the dissipation time which is the time scale after which a 2-point function of such operators decays to zero. As explained for instance in \cite{Maldacena:2015waa}, this means that in a large $N$ system with large $N$ factorization, after the dissipation time $F(t) \to F_d$ a time independent constant
\begin{equation}
F_d = \langle VV\rangle \langle WW\rangle ,
\end{equation}
as long as the times are independent of $N$.

In semiclassical systems, an exponential dependence on small changes, the butterfly effect, would translate into $[W(t),V] \propto i \hbar e^{\lambda _Lt} $ where $\lambda _L$ is the Lyapunov exponent, giving $C(t) \sim e^{2\lambda _Lt} $. One could define the Lyapunov exponent by such an exponential dependence of the double commutator. Correspondingly there should be an exponential dependence in the OTO correlator $F(t) = F_d - \epsilon \, e^{\lambda _Lt} +\cdots$ after the dissipation time (where a priori $\lambda _L$ here could be different). This gives another definition of the Lyapunov exponent as the leading exponent in the OTOC, which is the one used by \cite{Maldacena:2015waa}. Note that when $\epsilon $ is purely imaginary and the next exponential term in the expansion of $F$ with a doubled exponent has real coefficient, the two definitions coincide.

We also know that generally the scrambling time is logarithmic in a parameter which is the entropy in gravity. Therefore one could alternatively define the Lyapunov exponent by the coefficient of the logarithm in the scrambling time (as discussed in \cite{Aalsma:2020aib}). For instance, in gravity whether the first non-trivial dependence of the OTOC on time is $G_N e^{2\pi t/\beta } $ or $(G_N e^{2\pi t/\beta } )^2$, the times where this correction becomes of order one are the same and would lead to the same Lyapunov exponent by this specification. However, this cannot be used if we do not know a priori the analogue of $G_N$ in the system. In particular, we do not use this in our case.

The bound on chaos conjecture of \cite{Maldacena:2015waa} is that after the dissipation time a regularized version of the OTOC
\begin{equation}
\tilde F (t) = \tr [y V y W(t) y V y W(t) ],\qquad y = e^{-\beta H/4} / Z^{1/4} 
\end{equation}
satisfies
\begin{equation}
\frac{d}{dt} \big( F_d-\tilde F(t) \big) \le \frac{2\pi }{\beta } \big( F_d-\tilde F(t) \big)
\end{equation}
\emph{for real times}, which is usually stated as $\lambda _L \le \frac{2\pi }{\beta } $.

Having a doubled Lyapunov exponent might naively say that this immediately violates the bound on chaos, but let us try to be more specific.

In fact, the physical argument of \cite{Maldacena:2015waa} implicitly tells us something more. The basic argument considers the regularized OTOC on a strip in the complex $t$ plane of width $\beta /2$ in the imaginary direction. The maximum principle implies that in the interior of the strip the OTOC is bounded by the same value it is bounded by on its boundary. Therefore if $\tilde F(t) = F_d(1 - f e^{\lambda _Lt} +\cdots) $ we also infer that $f$ cannot be negative (since as shown in \cite{Maldacena:2015waa} this is the case on the boundary after enough time).

In our case the regularized version $\tilde F$ corresponds to $t_1=\frac{i}{4T} $, $t_2=t$, $t_3=-\frac{i}{4T} $, and $t_4=t-\frac{i}{2T} $. In the single copy this indeed results in a positive $f$. In the two copies, however, we get a \emph{negative} $f$ (in this case both $\ratiocoeff$ and $\ratiocoeff'$ have arguments $(-i\pi )$ and are thus positive, and there is an additional minus sign from the exponent). This does not happen when a bound on chaos holds. If we keep an overall small coefficient for the OTOC compared to the factorized result, one could naively argue that there is a priori no contradiction with the phrasing of the bound on chaos, but in any case if we allow choosing it differently as we saw in our system, we will get a result that does not satisfy the results of the bound on chaos. Naively, this is easiest to attribute to the choice of the time evolved operators we use in our setup.\footnote{In fact, these operators are remarked to be not `simple' operators, as required for the bound on chaos, already in the first footnote of \cite{Maldacena:2015waa}.} As mentioned, we could expect many issues to arise because of that, but we checked that they still satisfy many desired properties.

\section{Concluding remarks} \label{sec:remarks}

In this section we mainly make a few comments relating to other works.

The role of $G_N$ in the microscopic description is played by a semiclassical parameter that depends on the system we consider. For instance, in the Schwarzian theory there is a single dimensionful parameter, sometimes denoted by $C$. The semiclassical expansion in this case is when $C$ is large in units of the temperature, and the semiclassical parameter is $\frac{1}{CT} $. In ordinary SYK (not double-scaled) this parameter is essentially $\frac{\JMS}{TN} $ and so maps to large $N$. DSSYK has more parameters that change the qualitative behavior of the theory, where in addition to $N$, we have $\lambda $ as well as $\JMS$ in energy units. We saw that the holographic-semiclassical expansion corresponds to taking both small $\lambda $ and energy (above the ground state) much larger than $\JMS \lambda $, and the aforementioned semiclassical parameter is $\frac{\JMS \lambda }{T} $. If we work with $\JMS \sim O(1)$ this behaves as $\lambda $. More generally we can think about this as large temperature in appropriate units.

Recently, \cite{Kolchmeyer:2024fly}, related to \cite{Chandrasekaran:2022cip}, explicitly constructed a system of observers living in dS space, and also calculated the OTOC in this system. There are some curious similarities to the discussion here as well as differences. Even though two observers are considered, because of the isometry constraints that are relevant in this setup, the states are parametrized by one fluctuating energy variable, analogous to ours. In \cite{Kolchmeyer:2024fly} gravity is turned off, but still the OTOC has similar features to the gravitational effects, with the role of Newton's constant replaced by the mass of the observers. In our case we were not considering finite masses, and it would be interesting to understand how to incorporate this. This construction explicitly couples to quantum fields in the bulk of dS, which tells us correctly what a microscopic system is expected to give semiclassically.

Double-scaled SYK has played a crucial role in the relation to de Sitter studied in \cite{Narovlansky:2023lfz}. Indeed, this work used the maximal entropy state which corresponds to strictly infinite temperature. A doubled system was used in this work as well, although the details are slightly different, and \cite{Narovlansky:2023lfz} gauged the difference of the two Hamiltonians. One could wonder whether one of the two features, the infinite temperature state and the two copies of the system, was important for the relation to dS, or perhaps both. Our work suggests one possible answer to this question, as we use a particular doubled setup, but do not use the maximal entropy state.

A relation of DSSYK and dS space has also been suggested in \cite{Susskind:2021esx,Susskind:2022dfz,Lin:2022nss,Rahman:2022jsf,Susskind:2022bia}. In these papers a different time scale of DSSYK is proposed to be related to macroscopic dS physics. Related to this, there has been a question about the correct identification of the gravitational entropy parameter in the microscopic system \cite{Rahman:2023pgt}. From the present work point of view, the entropy can be studied using the OTOC. In this respect, it maps to a semiclassical parameter which depends on the particular system used, and our previous comment at the beginning of this section is relevant here. Note also that one should specify the ``good'' operators that describe the gravitational system, since the operators can also affect the OTOC behavior, as we saw. Indeed, both the operators and correspondingly the time scales relevant to the relation to dS discussed in these papers differ from the traditional ones in DSSYK, which can result in a different semiclassical parameter, as commented above. We have focuses on the traditional DSSYK regime since it is well understood and solved, but one can also study other interesting regimes. It would also be interesting to understand this discussion in terms of the dilaton-gravity and non-commutative geometry descriptions (see for instance \cite{Berkooz:2022mfk,Blommaert:2023opb,Blommaert:2023wad,Blommaert:2024ydx,Blommaert:2024whf,Belaey:2025ijg}), as well as other studied relations (as in \cite{Milekhin:2023bjv,Aguilar-Gutierrez:2024nau,Milekhin:2024vbb,Okuyama:2025hsd,Coleman:2021nor,Rabinovici:2023yex,Balasubramanian:2024lqk,Ambrosini:2024sre}).

\section*{Acknowledgments}

I would like to thank Dionysios Anninos, Patrick Draper, Thomas Faulkner, David Kolchmeyer, Jonah Kudler-Flam, Juan Maldacena, Alexey Milekhin, Robert Myers, Herman Verlinde, Jiuci Xu, and Zhenbin Yang for useful discussions.
VN gratefully acknowledges support from NSF Grant PHY-2207584.

\appendices

\section{Details of uncrossed correlators}

\subsection{2-point function} \label{sec:app_2pt_details}

In all correlation functions, because of energy conservation we can always add or subtract a fixed amount to all energies without changing the answer. We thus consider all energies as the energies above the ground state.

For a general operator of the form \eqref{eq:doubled_generic_ops}, applying the diagrammatic rules to the 2-point function \eqref{eq:2pf_diagram} we get
\begin{equation}
\begin{split}
& \int \frac{\lambda ^6 (q;q)_{\infty } ^6}{16\pi ^3 T \, \Gamma \left( \pm 2i(s_0+\omega )\right)^2 } d\omega  e^{-4\pi iT\omega t} \frac{\lambda ^{2(\Delta ^{(1)} +\Delta ^{(2)} )-6} }{(q;q)_{\infty } ^6 \Gamma (2\Delta ^{(1)} )\Gamma (2\Delta ^{(2)} )}\cdot \\
& \qquad \cdot \Gamma (\Delta ^{(1)}  \pm i\omega )\Gamma (\Delta ^{(2)}  \pm i\omega ) \Gamma \left( \Delta ^{(1)}  \pm i(2s_0+\omega )\right) \Gamma \left( \Delta ^{(2)}  \pm i(2s_0+\omega )\right) .
\end{split}
\end{equation}

One way to simplify this is to note that for large $M$ we have
\begin{equation} \label{eq:Gamma_large_arg}
\begin{split}
\Gamma (M+\alpha ) & \sim \Gamma (M) M^{\alpha } ,\\
\Gamma (-M+\alpha ) & \sim \Gamma (-M) M^{\alpha } e^{-i\pi \alpha } .
\end{split}
\end{equation}
The 2-point function then becomes
\begin{equation} \label{eq:2pf_general_op_dimensions}
\frac{(2\lambda s_0)^{2(\Delta ^{(1)} +\Delta ^{(2)} )} }{16 \pi ^3 T \, \Gamma (2\Delta ^{(1)} )\Gamma (2\Delta ^{(2)} )} \int d\omega \, e^{(2\pi -4\pi iTt)\omega } \Gamma (\Delta ^{(1)}  \pm i\omega ) \Gamma (\Delta ^{(2)}  \pm i\omega ).
\end{equation}

\subsection{Relation to the 2-point function in de Sitter} \label{sec:app_2pt_relation_dS}

Here we derive the relation of \eqref{eq:2pf_general_op_dimensions} to the 2-point function in de Sitter in general bulk dimension.

Let us start by evaluating
\begin{equation}
\Gamma (\Delta  \pm i\omega ) = \Gamma (2\Delta ) \int _0 ^1 du \, u^{\Delta +i\omega -1} (1-u)^{\Delta -i\omega -1} 
\end{equation}
which is an integral representation of the beta function. Calculating the Fourier integral of this, inserting $\int d\omega \, e^{i\omega t} $, the integral over $\omega $ gives a delta function localizing to $u=\frac{1}{1+e^t} $, so that
\begin{equation}
\int d\omega \, e^{i\omega t} \Gamma (\Delta \pm i\omega ) = \frac{\pi \, \Gamma (2\Delta )}{2^{ 2\Delta -1}\left( \cosh \frac{t}{2} \right) ^{2\Delta } } .
\end{equation}

Now we can calculate using convolution
\begin{equation}
\begin{split}
\int d\omega \, e^{i\omega t} & \Gamma (\Delta ^{(1)}  \pm i\omega )\Gamma (\Delta ^{(2)}  \pm i\omega ) =\\
&= \frac{\pi ^2\Gamma (2\Delta ^{(1)} )\Gamma (2\Delta ^{(2)} )}{2^{2(\Delta ^{(1)} +\Delta ^{(2)} )-2} } \int _{-\infty } ^{\infty } \frac{du}{2\pi } \left( \cosh \frac{u}{2} \right) ^{-2\Delta ^{(1)} } \left( \cosh \frac{t-u}{2} \right) ^{-2\Delta ^{(2)} } =\\
&=2\pi \Gamma (2\Delta ^{(1)} )\Gamma (2\Delta ^{(2)} ) \int _{-\infty } ^{\infty } du \, \frac{e^{u(\Delta ^{(1)} +\Delta ^{(2)} )-t\Delta ^{(2)} } }{(e^u+1)^{2\Delta ^{(1)} } (e^{u-t} +1 )^{2\Delta ^{(2)} } } =\\
& = 2\pi \Gamma (2\Delta ^{(1)} )\Gamma (2\Delta ^{(2)} )\int _0^{\infty } dy \, \frac{y^{\Delta ^{(1)} +\Delta ^{(2)} -1} e^{t\Delta ^{(2)} } }{(1+y)^{2\Delta ^{(1)} } (y+e^t)^{2\Delta ^{(2)} } } 
\end{split}
\end{equation}
where we changed variables to $e^u=y$. Now changing $y=\frac{w}{1-w} $ we get
\begin{equation}
\begin{split}
&\int d\omega \, e^{i\omega t} \Gamma (\Delta ^{(1)}  \pm i\omega )\Gamma (\Delta ^{(2)}  \pm i\omega ) = 2\pi \Gamma (2\Delta ^{(1)} )\Gamma (2\Delta ^{(2)} )e^{-t\Delta ^{(2)} } \\
& \qquad \qquad \cdot \int _0^1 dw \, w^{\Delta ^{(1)} +\Delta ^{(2)} -1} (1-w)^{\Delta ^{(1)} +\Delta ^{(2)} -1} \left( 1-w(1-e^{-t} )\right) ^{-2\Delta ^{(2)} } .
\end{split}
\end{equation}
The remaining integral is the Euler integral representation of the hypergeometric function, giving
\begin{equation}
\begin{split}
&\int d\omega \, e^{i\omega t} \Gamma (\Delta ^{(1)}  \pm i\omega )\Gamma (\Delta ^{(2)}  \pm i\omega ) = \frac{2\pi \Gamma (2\Delta ^{(1)} )\Gamma (2\Delta ^{(2)} )e^{-t\Delta ^{(2)} } \Gamma (\Delta ^{(1)} +\Delta ^{(2)} )^2}{\Gamma (2\Delta ^{(1)} +2\Delta ^{(2)} )} \\
& \qquad \qquad \cdot  {}_2F_1(2\Delta ^{(2)} ,\Delta ^{(1)} +\Delta ^{(2)};2(\Delta ^{(1)} +\Delta ^{(2)} );1-e^{-t} ).
\end{split}
\end{equation}
Using the transformations
\begin{equation}
\begin{split}
&{}_2F_1(a,b;2b;z)=(1-z)^{-a/2} \, {}_2F_1\left( \frac{a}{2} ,b-\frac{a}{2} ;b+\frac{1}{2} ;\frac{z^2}{4(z-1)} \right) \\
& {}_2F_1\left( a,b;a+b+\frac{1}{2} ;\sin^2 \phi \right) ={}_2F_1\left( 2a,2b;a+b+\frac{1}{2} ;\sin ^2\frac{\phi }{2} \right) 
\end{split}
\end{equation}
we can write the result also as
\begin{equation} \label{eq:Fourier_integral_giving_2F1}
\begin{split}
&\int d\omega \, e^{i\omega t} \Gamma (\Delta^{(1)}  \pm i\omega )\Gamma (\Delta ^{(2)}  \pm i\omega ) = \frac{2\pi \Gamma (2\Delta ^{(1)} )\Gamma (2\Delta ^{(2)} )\Gamma (\Delta ^{(1)} +\Delta ^{(2)} )^2}{\Gamma (2\Delta ^{(1)} +2\Delta ^{(2)} )} \\
& \qquad \qquad \cdot  {}_2F_1\left(2\Delta ^{(1)} ,2\Delta ^{(2)} ;\Delta ^{(1)} +\Delta ^{(2)} +\frac{1}{2};-\sinh^2 \frac{t}{4} \right).
\end{split}
\end{equation}
In order to have the Fourier transform in \eqref{eq:2pf_general_op_dimensions} we need to change here $t \to -4\pi Tt-2\pi i$ so that
\begin{equation}
\begin{split}
& \langle \cO (t) \cO (0)\rangle =\frac{(2\lambda s_0)^{2(\Delta ^{(1)} +\Delta ^{(2)} )} \Gamma (\Delta ^{(1)} +\Delta ^{(2)} )^2}{8\pi ^2T\, \Gamma (2\Delta ^{(1)} +2\Delta ^{(2)}) } \\
& \qquad \qquad \cdot {}_2F_1\left( 2\Delta ^{(1)} ,2\Delta ^{(2)} ;\Delta ^{(1)} +\Delta ^{(2)} +\frac{1}{2} ;\frac{1+\cosh (2\pi T t)}{2} \right) .
\end{split}
\end{equation}
We see that if we choose 
\begin{equation}
\Delta ^{(1)} +\Delta ^{(2)} =\frac{d-1}{2} 
\end{equation}
we get the 2-point function \eqref{eq:2pf_dS_2F1} of a scalar in $d$-dimensional de Sitter space. Parametrizing
\begin{equation}
2\Delta ^{(1),(2)} =\frac{d-1}{2} \pm i\mud 
\end{equation}
we get 
\begin{equation}
\begin{split}
& \langle \cO (t) \cO (0)\rangle = \frac{(2\lambda s_0)^{d-1}  \Gamma \left( \frac{d}{2} \right) \Gamma \left( \frac{d-1}{2} \right) ^2}{2\pi ^{d/2} T^{d-1} \, \Gamma (d-1)\Gamma (2\Delta ^{(1)} )\Gamma (2\Delta ^{(2)} )}  G_{d,\mud } (t)
\end{split}
\end{equation}
where $G_{d,\mud }$ is the 2-point function in $d$-dimensional de Sitter space \eqref{eq:2pf_dS_2F1}, with invariant distance $P=\cosh (2\pi Tt)$ (where $1/\rdS=2\pi T$). $\mud $ is related to the mass of the scalar by $\mud =\sqrt{m^2\rdS^2-\left( \frac{d-1}{2} \right) ^2}$. Note that again the microscopic temperature $T$ is identified with that in de Sitter $1/(2\pi \rdS)$, and the microscopic time is identified with the proper time in de Sitter.

\subsection{Uncrossed 4-point functions} \label{sec:app_uncrossed_4pt_details}

In this section we work with operators \eqref{eq:doubled_3d_ops}, so each operator is labeled by a single number $\Delta $. We also denote $t_{ij} =t_i-t_j$.
The diagram \eqref{eq:4pf_uncrossed_diagram_1} is simple and is just
\begin{equation}
\begin{split}
& \int \prod _{i=1} ^2 \left[ \frac{\lambda ^6 (q;q)_{\infty } ^6 d\omega _i}{16\pi ^3 T \, \Gamma \left( \pm 2i(s_0+\omega _i)\right) ^2} \right]  e^{-4\pi iT t_{21} \omega _1 - 4\pi iTt_{43} \omega _2} \frac{\lambda ^{-8}}{(q;q)_{\infty } ^{12} \Gamma (2\Delta _1)\Gamma (2-2\Delta _1) \Gamma (2\Delta _2)\Gamma (2-2\Delta _2)}
\cdot \\
& \cdot \Gamma (\Delta _1 \pm i\omega _1) \Gamma (1-\Delta _1 \pm i\omega _1) \Gamma \left( \Delta _1 \pm i(2s_0+\omega _1)\right) \Gamma \left(1- \Delta _1 \pm i(2s_0+\omega _1)\right) \cdot \\
& \cdot \Gamma (\Delta _2 \pm i\omega _2) \Gamma (1-\Delta _2 \pm i\omega _2) \Gamma \left( \Delta _2 \pm i(2s_0+\omega _2)\right) \Gamma \left(1- \Delta _2 \pm i(2s_0+\omega _2)\right) .
\end{split}
\end{equation}
This clearly factorizes, with each factor being the same as the 2-point function.

The diagram \eqref{eq:4pf_uncrossed_diagram_2} instead is given by
\begin{equation}
\begin{split}
& \int \prod _{i=1} ^2 \left[ \frac{\lambda ^6 (q;q)_{\infty } ^6 d\omega _i}{16\pi ^3 T } \right]  \frac{e^{-4\pi iT \omega _1(t_{43} +t_{21} ) - 4\pi iTt_{32} (\omega _1+\omega _2)}}{\Gamma \left( \pm 2i(s_0+\omega _1)\right) ^2 \Gamma \left( \pm 2i(s_0+\omega _1+\omega _2)\right) ^2} \cdot \\
& \cdot \frac{\lambda ^{-8}}{(q;q)_{\infty } ^{12} \Gamma (2\Delta _1)\Gamma (2-2\Delta _1) \Gamma (2\Delta _2)\Gamma (2-2\Delta _2)}
\cdot \\
& \cdot \Gamma (\Delta _1 \pm i\omega _1) \Gamma (1-\Delta _1 \pm i\omega _1) \Gamma \left( \Delta _1 \pm i(2s_0+\omega _1)\right) \Gamma \left(1- \Delta _1 \pm i(2s_0+\omega _1)\right) \cdot \\
& \cdot \Gamma (\Delta _2 \pm i\omega _2) \Gamma (1-\Delta _2 \pm i\omega _2) \Gamma \left( \Delta _2 \pm i(2s_0+2\omega _1+\omega _2)\right) \Gamma \left(1- \Delta _2 \pm i(2s_0+2\omega _1+\omega _2)\right) .
\end{split}
\end{equation}
This does not manifestly factorize, but in the holographic-semiclassical limit we get
\begin{equation}
\begin{split}
& \frac{(2s_0\lambda )^4}{(16\pi ^3T)^2 \Gamma (2\Delta _1)\Gamma (2-2\Delta _1)\Gamma (2\Delta _2)\Gamma (2-2\Delta _2)} \cdot \\
& \cdot \int d\omega _1 \exp \left[ (2\pi -4\pi iT t_{41} )\omega _1 \right]
\Gamma (\Delta _1 \pm i\omega _1) \Gamma (1-\Delta _1 \pm i\omega _1) 
\cdot \\
& \cdot 
\int d\omega _2 \exp \left[  (2\pi -4\pi iT t_{32} )\omega _2\right]
\Gamma (\Delta _2 \pm i\omega _2) \Gamma (1-\Delta _2 \pm i\omega _2) .
\end{split}
\end{equation}

\section{The one copy system} \label{sec:app_one_copy_crossed_4pf}

First we notice that for a single copy, the 2-point function is
\begin{equation} \label{eq:one_copy_2pf_details}
\begin{split}
& \int \frac{\lambda ^3 (q;q)_{\infty } ^3}{2\pi \Gamma \left( \pm 2i(s _0+\omega )\right) } d\omega \, e^{-2\pi iT_1\omega t} \frac{\lambda ^{2\Delta -3} }{(q;q)_{\infty } ^3\Gamma  (2\Delta )} \Gamma (\Delta \pm i\omega )\Gamma \left( \Delta  \pm i(2s_0+\omega )\right) \\
& \sim (2\lambda s_0)^{2\Delta } \int \frac{d\omega }{2\pi } e^{\pi \omega -2\pi iT_1\omega t} \frac{\Gamma (\Delta \pm i\omega )}{\Gamma (2\Delta )} 
 = (2\lambda s_0)^{2\Delta } \left( 2\sin(i\pi T_1 t)\right) ^{-2\Delta } .
\end{split}
\end{equation}
We have denoted the temperature of a single copy by $T_1$.

Next we move on to our main interest which is the crossed 4-point function. For a one copy system this essentially was studied in \cite{Berkooz:2018jqr}, and can be thought of as a special case of the two copies calculation we do in App.\ \ref{sec:app_crossed_4pf_details}.
Instead of the expression \eqref{eq:crossed_4pf_int_expr} there, we get
\begin{equation} \label{eq:crossed_4pf_one_copy_int_expr}
\begin{split}
& (2\lambda s_0)^{2\Delta _1+2\Delta _2}
\Big[ \Gamma (2\Delta _1) \Gamma (2\Delta _2)\Big]^{-1} (2\pi )^{-3} \\
& \int \prod _{i=1} ^3 \left[ d\omega _i \right]
\exp \bigg[ \Big(\frac{\pi}{2} -2\pi iT_1 t_{21} \Big)\omega _1 +\Big(\frac{\pi}{2} -2\pi iT_1 t_{32} \Big)\omega _2 +\Big(\frac{\pi}{2} -2\pi iT_1 t_{43} \Big)\omega _3 \bigg] \\
\bigg\{
& \Gamma (\Delta _1+i\omega _{23} )\Gamma (\Delta _1-i\omega _1)\Gamma (\Delta _2+i\omega _{21} )\Gamma (\Delta _2-i\omega _3)\Gamma \left( i(\omega _1+\omega _3-\omega _2)\right) (2s_0)^{i(\omega _1-\omega _2+\omega _3)} \\
& + \Gamma (\Delta _1+i\omega _{32} )\Gamma (\Delta _1+i\omega _1)\Gamma (\Delta _2+i\omega _{12} )\Gamma (\Delta _2+i\omega _3)\Gamma \left( i(\omega _2-\omega _1-\omega _3)\right) (2s_0)^{i(-\omega _1+\omega _2-\omega _3)} \bigg\} .
\end{split}
\end{equation}

\subsection{Exact evaluation}

In this case, the integrals can be calculated exactly. For the first term, coming from picking the first summand in the curly brackets, we obtained
\begin{equation} \label{eq:crossed_4pf_one_copy_exact_first_term}
\begin{split}
& (2\lambda s_0)^{2\Delta _1+2\Delta _2} \left( 2\sin(i\pi T_1t_{31} )\right) ^{-2\Delta _1} \left( 2\sin(i\pi T_1t_{42} )\right) ^{-2\Delta _2} x^{2\Delta _1} U(2\Delta _1,1+2\Delta _1-2\Delta _2,x)
\end{split}
\end{equation}
where
\begin{equation}
x=8s_0i \sinh(\pi T_1t_{31} )\sinh(\pi T_1t_{42} )e^{\pi T_1 (t_1+t_3-t_2-t_4)} .
\end{equation}
The function $U$ is a confluent hypergeometric function, a combination of ${}_1F_1$ functions, and is explicitly given by
\begin{equation} \label{eq:U_func_def}
U(\alpha ,\gamma ;z)=\frac{1}{\Gamma (\alpha )} \int _0^{\infty } dt \, e^{-zt} t^{\alpha -1} (t+1)^{\gamma -\alpha -1} .
\end{equation}
It is also denoted sometimes by $\Psi (\alpha ,\gamma ;z)$.

The chaotic OTO regime is when $t_2,t_4 \sim t$ and $t_1,t_3 \sim 0$ and $t$ is large. We can take $t$ large and positive or $t$ large and negative, so without loss of generality we take $t$ positive.
In order to probe the scrambling time, we take $s_0$ and $t$ large such that $x$ is kept finite.

The second term in \eqref{eq:crossed_4pf_one_copy_int_expr} is obtained by replacing $t_i \to -t^*_i$ and taking complex conjugation of the entire expression. Depending on the sign of $t$, the dominant time dependence comes from one of the two terms. Since we have chosen positive $t$, the important piece is the first term and so we concentrate on it, while the second term simply vanishes in the semiclassical limit for any time in this regime.\footnote{In the discussion in this subsection, we consider a contour for the frequencies along $\mathbb{R} -i\epsilon $.}

We can normalize the 4-point function by a product of 2-point functions which takes a simpler form in the regime of large positive $t$
\begin{equation} \label{eq:one_copy_OTO_result}
\begin{split}
& \frac{\langle \cO _2(t_4) \cO _1(t_3) \cO _2(t_2) \cO _1(t_1) \rangle\onecopy}{\langle \cO _2(t_4) \cO _2(t_2)\rangle\onecopy \langle \cO _1(t_3) \cO _1(t_1)\rangle\onecopy } = x^{2\Delta _1} U(2\Delta _1,1+2\Delta _1-2\Delta _2,x).
\end{split}
\end{equation}

In the chaotic regime, taking $t \to \infty $ much later than the scrambling time, we have $x \to 0$. In this limit
\begin{equation}
U(\alpha ,\gamma ;x) = \frac{\Gamma (\gamma -1)}{\Gamma (\alpha )} x^{1-\gamma } +\cdots 
\end{equation}
plus terms that are either smaller or remain finite; the quoted term gives that \eqref{eq:one_copy_OTO_result} behaves as $x^{2\Delta_2} $ and goes to zero, and the latter clearly give the same result. The vanishing of the 4-point function in this limit is what one expects.

The chaotic behavior is associated with the early time onset of building a non-zero value for the double-commutator. This corresponds to the same chaotic regime discussed before, but when the time is much earlier than the scrambling time. This translates to fixed $x$ but such that $|x| \gg 1$. In this case we can estimate the asymptotic behavior using
\begin{equation}
x^{\alpha } U(\alpha ,\gamma ;x) \sim {}_2F_0(\alpha ,\alpha -\gamma +1; -x^{-1} ) = 1-\alpha (\alpha -\gamma +1) x^{-1} + \cdots 
\end{equation}
giving
\begin{equation} \label{eq:app_one_copy_OTO_expansion}
\frac{\langle \cO _2(t_4) \cO _1(t_3) \cO _2(t_2) \cO _1(t_1) \rangle\onecopy}{\langle \cO _2(t_4) \cO _2(t_2)\rangle\onecopy \langle \cO _1(t_3) \cO _1(t_1)\rangle\onecopy } \sim 1+ \frac{i}{2} \frac{\Delta _1\Delta _2}{s_0 \sinh(\pi T_1 t_{31} ) \sinh(\pi T_1 t_{42} )} e^{\pi T_1(t_2+t_4-t_1-t_3)} + \cdots .
\end{equation}

\subsection{Perturbative evaluation} \label{sec:app_one_copy_pert}

Even though we could evaluate the correlator in this case, as a warmup for the two-copy system let us obtain the expansion \eqref{eq:app_one_copy_OTO_expansion} in a different way which does not require knowing the full answer (see \cite{Kolchmeyer:2024fly}).

For any finite times, in the semiclassical limit the 4-point function just factorizes into a product of 2-point functions. This does not happen for times around the scrambling time.

\textbf{First term}. Let us start with the first term in \eqref{eq:crossed_4pf_one_copy_int_expr} without the coefficient in the first line
\begin{equation} \label{eq:one_copy_crossed_pert_first_term}
\begin{split}
& \int \prod _{i=1} ^3 \left[ d\omega _i \right]
\exp \bigg[ \Big(\frac{\pi}{2} +i \log(2s_0)-2\pi iT_1 t_{21} \Big)\omega _1 +\Big(\frac{\pi}{2} -i \log(2s_0) -2\pi iT_1 t_{32} \Big)\omega _2 \\
& \qquad \qquad +\Big(\frac{\pi}{2}+i \log(2s_0) -2\pi iT_1 t_{43} \Big)\omega _3 \bigg] \\
& \Gamma (\Delta _1+i\omega _{23} )\Gamma (\Delta _1-i\omega _1)\Gamma (\Delta _2+i\omega _{21} )\Gamma (\Delta _2-i\omega _3)\Gamma \left( i(\omega _1+\omega _3-\omega _2)\right) .
\end{split}
\end{equation}
Indeed we see that for finite times there are strong oscillations from the $s_0$ terms that would result in a $\delta (\omega _1-\omega _2+\omega _3)$. In order to emphasize the scrambling time, let us temporarily use the following redundant parametrization (in terms of new variables $\td_i$ and one of $t$ and $u$)
\begin{equation}
\begin{split}
& t_2= t+\td_2,\quad t_4=t+\td_4 ,\qquad t=\frac{1}{2\pi T_1} \log (2s_0) + u, \\
& t_1 = \td_1, \quad t_3= \td_3 .
\end{split}
\end{equation}
In addition, we change variables to
\begin{equation} \label{eq:crossed_4pf_pert_evaluation_energies_cov}
\Omega =\omega _1-\omega _2+\omega _3,\quad \Omega _1=\omega _1,\quad \Omega _2=\omega _2-\omega _1
\end{equation}
or
\begin{equation}
\omega _1=\Omega _1,\quad \omega _2=\Omega _1+\Omega _2,\quad \omega _3=\Omega +\Omega _2.
\end{equation}
With these definitions, \eqref{eq:one_copy_crossed_pert_first_term} becomes
\begin{equation}
\begin{split}
& \int d\Omega d\Omega _1d\Omega _2 \exp \bigg[ \Big( \frac{\pi }{2} -2\pi iT_1u-2\pi iT_1 \td_{43} \Big)\Omega  + \Big( \pi -2\pi iT_1 \td_{31} \Big) \Omega _1 + \Big(\pi -2\pi iT_1 \td_{42} \Big) \Omega _2 \bigg] \\
& \Gamma (\Delta _1+i\Omega _1-i\Omega )\Gamma (\Delta _1-i\Omega _1)\Gamma (\Delta _2+i\Omega _2)\Gamma (\Delta _2-i\Omega _2-i\Omega )\Gamma (i\Omega ) .
\end{split}
\end{equation}

For the chaotic behavior, we are interested in $u \to -\infty $, meaning that we should close the $\Omega $ contour above. Assuming $\Delta _1,\Delta _2>0$, all Gamma functions with argument $(-i\Omega )$ have poles below the real axis, and have $\Delta _i$-dependence. Above or at the real axis we have poles only from $\Gamma (i\Omega )$ at $\Omega =in$ with $n=0,1,2,\cdots $. These poles give $u$-dependent exponents $e^{2\pi T_1nu} $ which become smaller and smaller corrections for larger $n$. We will be interested in the leading $n=0,1$ contributions.\footnote{In this subsection, we use the symmetric principal value prescription which averages the $\mathbb{R} +i\epsilon $ and $\mathbb{R} -i\epsilon $ contours, each weighted by a $1/2$. However, we emphasize that this is just one possible choice.} We find
\begin{equation}
\begin{split}
& \pi \int d\Omega _1 \, e^{(\pi -2\pi iT_1 \td_{31} )\Omega _1} \Gamma (\Delta _1 \pm i\Omega _1) \int d\Omega _2 \, e^{(\pi -2\pi iT_1 \td_{42} )\Omega _2} \Gamma (\Delta _2 \pm i\Omega _2) \\
& -2\pi \frac{i}{2s_0} e^{2\pi T_1t+2\pi T_1 \td_{43} }
\int d\Omega _1 \, e^{(\pi -2\pi iT_1 \td_{31} )\Omega _1} (\Delta _1+i\Omega _1)\Gamma (\Delta _1 \pm i\Omega _1) \\
& \qquad \qquad \qquad \qquad  \int d\Omega _2 \, e^{(\pi -2\pi iT_1 \td_{42} )\Omega _2} (\Delta _2-i\Omega _2)\Gamma (\Delta _2 \pm i\Omega _2) +\cdots .
\end{split}
\end{equation}

Instead, at very long times $u \to \infty $. In this case we close the contour below, getting contribution from various $\Delta _i$-dependent poles. However, we will only be interested in the leading term coming from the $\Omega =0$ pole, with the subleading terms going to zero. We get immediately
\begin{equation}
\begin{split}
& - \pi \int d\Omega _1 \, e^{(\pi -2\pi iT_1 \td_{31} )\Omega _1} \Gamma (\Delta _1 \pm i\Omega _1) \int d\Omega _2 \, e^{(\pi -2\pi iT_1 \td_{42} )\Omega _2} \Gamma (\Delta _2 \pm i\Omega _2) .
\end{split}
\end{equation}

\textbf{Second term}. 
The second term in \eqref{eq:crossed_4pf_one_copy_int_expr} without the coefficient in the first line is
\begin{equation} \label{eq:one_copy_crossed_pert_second_term}
\begin{split}
& \int \prod _{i=1} ^3 \left[ d\omega _i \right]
\exp \bigg[ \Big(\frac{\pi}{2} -i \log(2s_0)-2\pi iT_1 t_{21} \Big)\omega _1 +\Big(\frac{\pi}{2} +i \log(2s_0) -2\pi iT_1 t_{32} \Big)\omega _2 \\
& \qquad \qquad +\Big(\frac{\pi}{2}-i \log(2s_0) -2\pi iT_1 t_{43} \Big)\omega _3 \bigg] \\
& \Gamma (\Delta _1+i\omega _{32} )\Gamma (\Delta _1+i\omega _1)\Gamma (\Delta _2+i\omega _{12} )\Gamma (\Delta _2+i\omega _3)\Gamma \left( i(\omega _2-\omega _1-\omega _3)\right) .
\end{split}
\end{equation}
Doing the same change of variables we have
\begin{equation}
\begin{split}
& \int d\Omega d\Omega _1d\Omega _2 \\
& \exp \bigg[ \Big( \frac{\pi }{2} -2i\log(2s_0)-2\pi iT_1u-2\pi iT_1 \td_{43} \Big)\Omega  + \Big( \pi -2\pi iT_1 \td_{31} \Big) \Omega _1 + \Big(\pi -2\pi iT_1 \td_{42} \Big) \Omega _2 \bigg] \\
& \Gamma (\Delta _1-i\Omega _1+i\Omega )\Gamma (\Delta _1+i\Omega _1)\Gamma (\Delta _2-i\Omega _2)\Gamma (\Delta _2+i\Omega _2+i\Omega )\Gamma (-i\Omega ) .
\end{split}
\end{equation}
We see that in this time regime we localize around $\Omega =0$. Around this
\begin{equation}
\Gamma (-i\Omega )e^{-2i\log(2s_0)\Omega } \sim i \frac{\cos(2\log(2s_0)\Omega )}{\Omega } + \frac{\sin(2\log(2s_0)\Omega )}{\Omega }  \to \pi \delta (\Omega )
\end{equation}
since in our conventions
\begin{equation} \label{eq:lim_of_cos_sin}
\frac{\cos(M\Omega )}{\Omega }  \xrightarrow{M \to \infty } 0,\qquad \frac{\sin(M\Omega )}{\Omega } \xrightarrow{M \to \infty } \pi \delta (\Omega ).
\end{equation}
Therefore, independently of what $u$ we take, the second term just gives
\begin{equation}
\begin{split}
&  \pi \int d\Omega _1 \, e^{(\pi -2\pi iT_1 \td_{31} )\Omega _1} \Gamma (\Delta _1 \pm i\Omega _1) \int d\Omega _2 \, e^{(\pi -2\pi iT_1 \td_{42} )\Omega _2} \Gamma (\Delta _2 \pm i\Omega _2) .
\end{split}
\end{equation}

\textbf{Putting the two terms together}. For $u \to \infty $ we get a cancellation of the two terms. This is consistent with the exact evaluation and with what we expect. For the chaotic regime $u \to -\infty $ instead we get the expansion
\begin{equation}
\begin{split}
& 2\pi \int d\Omega _1 \, e^{(\pi -2\pi iT_1 \td_{31} )\Omega _1} \Gamma (\Delta _1 \pm i\Omega _1) \int d\Omega _2 \, e^{(\pi -2\pi iT_1 \td_{42} )\Omega _2} \Gamma (\Delta _2 \pm i\Omega _2) \\
& -2\pi \frac{i}{2s_0} e^{2\pi T_1t+2\pi T_1 \td_{43} }
\int d\Omega _1 \, e^{(\pi -2\pi iT_1 \td_{31} )\Omega _1} (\Delta _1+i\Omega _1)\Gamma (\Delta _1 \pm i\Omega _1) \\
& \qquad \qquad \qquad \qquad  \int d\Omega _2 \, e^{(\pi -2\pi iT_1 \td_{42} )\Omega _2} (\Delta _2-i\Omega _2)\Gamma (\Delta _2 \pm i\Omega _2) +\cdots .
\end{split}
\end{equation}
When restoring the prefactors, this reproduces the expansion of \eqref{eq:crossed_4pf_one_copy_exact_first_term}. For instance, at leading order $x^{2\Delta _1} U( \cdots )$ is 1, and we remain in \eqref{eq:crossed_4pf_one_copy_exact_first_term} with the product of 2-point functions. The leading behavior here is also the product of the integral representation of the 2-point functions we saw in \eqref{eq:one_copy_2pf_details}.

\section{Crossed 4-point function calculation} \label{sec:app_crossed_4pf_details}

The derivation of the crossed 4-point function follows the logic we discussed in setting up the system, with projected states in the two copies of the correlation functions. For each copy, there is a new ingredient to the diagrammatic rules which is that the crossing of chords in the bulk of the disc is assigned a $6j$ symbol of a $q$-deformation of $\mathfrak{sl}_2$ \cite{Berkooz:2018jqr}. In the following we first write the full expressions without taking any limits, except for taking small $\lambda $ in terms that do not include any energies. In order words, we use small $\lambda $, but do not assume anything about the energies. We can write the resulting expression as
\begin{equation} \label{eq:crossed_1}
\begin{split}
& \int \prod _{i=1} ^3 \left[ \frac{\lambda ^6(q;q)_{\infty } ^6}{16\pi ^3 T\, \Gamma _q(\pm 2is_i) ^2} d\omega _i\right]  e^{-4\pi iT ( \omega _1 t_{21} +\omega _2 t_{32} +\omega _3 t_{43} )}
\bigg\{ (1-q)^{-9+2\Delta _1+2\Delta _2} (q;q)_{\infty } ^{-9} \\
& \frac{\Gamma _q(\Delta _1+2\Delta _2+is_{2+3} )\Gamma _q(\Delta _1+is_{\pm 2 \pm 3} )\Gamma _q(\Delta _1+is_{\pm 1 \pm 4} )\Gamma _q(\Delta _2+is_{\pm 1 \pm 2} )\Gamma _q(\Delta _2+is _{\pm 3 \pm 4} )}{\Gamma _q(2\Delta _1)\Gamma _q(2\Delta _2)^2\Gamma _q(\Delta _1-is_{2+3} )\Gamma _q(\Delta _1+\Delta _2+is_{3 \pm 1} )\Gamma _q(\Delta _1+\Delta _2+is_{2 \pm 4} )} \\
& {}_8W_7 \Big( q^{\Delta _1+2\Delta _2-1+is_{2+3}} ;q^{\Delta _1+is_{2+3} } ,q^{\Delta _2+is_{2 \pm 1} } ,q^{\Delta _2+is_{3 \pm 4} } ;q,q^{\Delta _1+is _{-2-3} } \Big)  \bigg\} \cdot (\Delta _i \to 1-\Delta _i) .
\end{split}
\end{equation}
Let us explain the notation used. First, temporarily we included $s_4$ which in our case will become $s_0$. Second, $s_i = s_0+\omega _i$ for $i=1,2,3$. Addition in the subscript of $s$ means the addition of the corresponding $s$ variables, that is $s_{\pm i \pm j \pm \cdots } =\pm s_i \pm s_j \pm \cdots $. For Gamma functions with multiple arguments coming from multiple options for addition of $s$ variables, we include a product of all possible Gamma functions; for example, $\Gamma (is_{1 \pm 2} )=\Gamma (is_{1+2} )\Gamma (is_{1-2} )$. $\Gamma _q$ is the $q$-Gamma function. As before, $t_{ij} =t_i-t_j$. The function ${}_8W_7$ is an instance of a very-well-poised basic hypergeometric series. It is given by
\begin{equation}
{}_8W_7(A;a,b,c,d,e;q,z)={}_8\phi _7 \left[ \begin{matrix}
A,qA^{\frac{1}{2}} ,-qA^{\frac{1}{2}} ,a,b,c,d,e \\
A^{\frac{1}{2}  } ,-A^{\frac{1}{2} } ,\frac{qA}{a} ,\cdots \frac{qA}{e} 
\end{matrix};q,z \right] 
\end{equation}
where $\phi $ is the basic hypergeometric series. We are concerned with very-well-poised basic hypergeometric series satisfying the balancing condition $abcdez=A^2 q^2$. The function ${}_8W_7$ has various symmetry properties, including invariance under permutations of the $a,b,c,d,e$ arguments, manifest in its definition. It also has a symmetry property implying that
\begin{equation}
\begin{split}
& {}_8W_7 \Big( q^{\Delta _1+2\Delta _2-1+is_{2+3}} ;q^{\Delta _1+is_{2+3} } ,q^{\Delta _2+is_{2 \pm 1} } ,q^{\Delta _2+is_{3 \pm 4} } ;q,q^{\Delta _1+is _{-2-3} } \Big) = \\
&= \frac{\Gamma _q(\Delta _1+\Delta _2+is_{3 \pm 1} )\Gamma _q(\Delta _1+\Delta _2+is_{2 \pm 4} )\Gamma _q(\Delta _1+2\Delta _2+is_{-2-3} )\Gamma _q(\Delta _1+is_{-2-3} )}{\Gamma _q(\Delta _1+2\Delta _2+is_{2+3} )\Gamma _q(\Delta _1+is_{2+3} )\Gamma _q(\Delta _1+\Delta _2+is_{ \pm 1-3} )\Gamma _q(\Delta _1+\Delta _2+is_{\pm 4-2} )} \cdot \\
& \cdot {}_8W_7 \Big( q^{\Delta _1+2\Delta _2-1+is_{-2-3} } ;q^{\Delta _2+is_{1-2} } ,q^{\Delta _2+is_{-3 \mp 4}} q^{\Delta _2+is_{-1-2} } ,q^{\Delta _1+is_{-2-3} } ;q,q^{\Delta _1+is_{2+3} } \Big) .
\end{split}
\end{equation}
For more details see \cite{Berkooz:2018jqr} (in particular, equation (6.1)). Using this identity, we can rewrite \eqref{eq:crossed_1} as
\begin{equation}
\begin{split}
&\lambda ^{2\Delta _1+2(1-\Delta _1)+2\Delta _2+2(1-\Delta _2)} \int \prod _{i=1} ^3 \left[ \frac{d\omega _i}{16\pi ^3T \, \Gamma _q(\pm 2is_i)^2} \right] 
e^{-4\pi iT ( \omega _1 t_{21} +\omega _2 t_{32} +\omega _3 t_{43} )} 
 \bigg\{ \Gamma _q(\Delta _1+2\Delta _2+is _{-2-3} )\\
& \frac{\Gamma _q(\Delta _1+is_{-2 \pm 3} )\Gamma _q(\Delta _1+is_{2-3} )\Gamma _q(\Delta _1+is_{\pm 1 \pm 4} )\Gamma _q(\Delta _2+is_{\pm 1 \pm 2} )\Gamma _q(\Delta _2+is_{\pm 3 \pm 4} )}{\Gamma _q(2\Delta _1)\Gamma _q(2\Delta _2)^2 \Gamma _q(\Delta _1+\Delta _2+is_{\pm 1 -3} )\Gamma _q(\Delta _1+\Delta _2+is_{\pm 4-2} )} \\
& {}_8W_7 \Big( q^{\Delta _1+2\Delta _2-1+is_{-2-3} } ;q^{\Delta _2+is_{1-2} } ,q^{\Delta _2+is_{-3 \mp 4}} q^{\Delta _2+is_{-1-2} } ,q^{\Delta _1+is_{-2-3} } ;q,q^{\Delta _1+is_{2+3} } \Big) \bigg\} \cdot(\Delta _i \to 1-\Delta _i) .
\end{split}
\end{equation}

Now we restrict to the holographic-semiclassical limit and use the result \eqref{eq:lim_of_8W7} we derive in App.\ \ref{sec:limit_of_W} to get a simpler expression
\begin{equation}
\begin{split}
&\lambda ^{2\Delta _1+2(1-\Delta _1)+2\Delta _2+2(1-\Delta _2)} \int \prod _{i=1} ^3 \left[ \frac{d\omega _i}{16\pi ^3T \, \Gamma (\pm 2is_i)^2} \right] 
e^{-4\pi iT ( \omega _1 t_{21} +\omega _2 t_{32} +\omega _3 t_{43} )} \\
&
\Big[ \Gamma (2\Delta _1) \Gamma (2\Delta _2)\Gamma (2-2\Delta _1)\Gamma (2-2\Delta _2)\Big]^{-1}   \bigg\{ \\
&   \Gamma (\Delta _1+is_{-3 \pm 2} ) \Gamma (\Delta _1+is_{4 \pm 1} )\Gamma (\Delta _2+is_{-1 \pm 2} )\Gamma (\Delta _2+is_{4 \pm 3} )\Gamma (-2is_4) 
 \Gamma (is_{1+3-2-4} )\Gamma (is_{1+2+3-4} ) \\
& + \Gamma (\Delta _1+is_{-2 \pm 3} )\Gamma (\Delta _1+is_{1 \pm 4} )\Gamma (\Delta _2+is_{-2 \pm 1} )\Gamma (\Delta _2+is_{3 \pm 4} )\Gamma (2is_2)
\Gamma (is_{2+4-1-3} )\Gamma (is_{2-1-3-4} ) \\
& \bigg\} \cdot(\Delta _i \to 1-\Delta _i) .
\end{split}
\end{equation}
Using \eqref{eq:Gamma_large_arg} this further simplifies to (where we denote $\omega _{ij} =\omega _i-\omega _j$)
\begin{equation} \label{eq:crossed_4pf_int_expr_before_simplify}
\begin{split}
& \langle \cO _2(t_4) \cO _1(t_3) \cO _2(t_2) \cO _1(t_1) \rangle = \\
& = (2\lambda s_0)^{2\Delta _1+2(1-\Delta _1)+2\Delta _2+2(1-\Delta _2)}
\Big[ \Gamma (2\Delta _1) \Gamma (2\Delta _2)\Gamma (2-2\Delta _1)\Gamma (2-2\Delta _2)\Big]^{-1} \big(16 \pi ^3T\big)^{-3} \\
& \int \prod _{i=1} ^3 \left[ d\omega _i  \right]
\exp \Big[ (\pi -4\pi iT t_{21} )\omega _1 +(\pi -4\pi iT t_{32} )\omega _2 +(\pi -4\pi iT t_{43} )\omega _3 \Big] \\
\bigg\{
& \Gamma (\Delta _1+i\omega _{23} )\Gamma (\Delta _1-i\omega _1)\Gamma (\Delta _2+i\omega _{21} )\Gamma (\Delta _2-i\omega _3)\Gamma \left( i(\omega _1+\omega _3-\omega _2)\right) (2s_0)^{i(\omega _1-\omega _2+\omega _3)} \\
& + \Gamma (\Delta _1+i\omega _{32} )\Gamma (\Delta _1+i\omega _1)\Gamma (\Delta _2+i\omega _{12} )\Gamma (\Delta _2+i\omega _3)\Gamma \left( i(\omega _2-\omega _1-\omega _3)\right) (2s_0)^{i(-\omega _1+\omega _2-\omega _3)} \bigg\} \\
& \cdot (\Delta _i \to 1-\Delta _i) .
\end{split}
\end{equation}
This is the expression we work with. One expects that at finite times the 4-point function factorizes. In fact, this is non-trivial in this setup. We analyze this in Sec.\ \ref{sec:finite_times}, but first we consider times of the order of the scrambling time.

\subsection{Perturbative evaluation} \label{sec:crossed_pert_evaluation}

In this case we have not evaluated the integral to get a closed form expression. Instead we proceed with a perturbative evaluation similar to App.\ \ref{sec:app_one_copy_pert}. When opening the curly brackets in \eqref{eq:crossed_4pf_int_expr_before_simplify} we get four terms. In all terms below we at first do not write the coefficient appearing in the first line of \eqref{eq:crossed_4pf_int_expr_before_simplify}.

\textbf{First term}. First we have the term
\begin{equation} \label{eq:crossed_4pf_first_term}
\begin{split}
& \int \prod _{i=1} ^3 \left[ d\omega _i  \right]
\exp \Big[ (\pi +2i\log(2s_0)-4\pi iT t_{21} )\omega _1 +(\pi -2i\log(2s_0)-4\pi iT t_{32} )\omega _2 \\
& \qquad \qquad +(\pi +2i\log(2s_0)-4\pi iT t_{43} )\omega _3 \Big] \\
& \Gamma (\Delta _1+i\omega _{23} )\Gamma (\Delta _1-i\omega _1)\Gamma (\Delta _2+i\omega _{21} )\Gamma (\Delta _2-i\omega _3)\Gamma \left( i(\omega _1+\omega _3-\omega _2)\right)^2 \\
& \Gamma (1-\Delta _1+i\omega _{23} )\Gamma (1-\Delta _1-i\omega _1)\Gamma (1-\Delta _2+i\omega _{21} )\Gamma (1-\Delta _2-i\omega _3) .
\end{split}
\end{equation}
Again we are interested in times of the order of the scrambling time, and positive, in which case this term has a non-trivial functional form.
We use the parametrization
\begin{equation} \label{eq:crossed_scrambling_time_cov}
\begin{split}
& t_2= t+\td_2,\quad t_4=t+\td_4 ,\qquad t=\frac{1}{2\pi T} \log (2s_0) + u, \\
& t_1 = \td_1, \quad t_3= \td_3 ,
\end{split}
\end{equation}
and change variables for the frequencies to \eqref{eq:crossed_4pf_pert_evaluation_energies_cov}. This gives
\begin{equation} \label{eq:crossed_4pf_first_term_after_cov}
\begin{split}
& \int d\Omega d\Omega _1d\Omega _2 \, 
\exp \Big[ \big(\pi-4\pi iTu-4\pi iT \td_{43} \big)\Omega  +\big(2\pi -4\pi iT \td_{31} \big)\Omega  _1 +\big(2\pi-4\pi iT \td_{42} \big)\Omega _2 \Big] \\
& \Gamma (\Delta _1+i\Omega _1-i\Omega  )\Gamma (\Delta _1-i\Omega _1)\Gamma (\Delta _2+i\Omega _2 )\Gamma (\Delta _2-i\Omega _2-i\Omega )\Gamma \left( i\Omega \right)^2 \\
& \Gamma (1-\Delta _1+i\Omega _1-i\Omega  )\Gamma (1-\Delta _1-i\Omega _1)\Gamma (1-\Delta _2+i\Omega _2 )\Gamma (1-\Delta _2-i\Omega _2-i\Omega ).
\end{split}
\end{equation}
In our construction, from the SYK point of view, we should consider $0<\Delta _i<1$, and so all Gamma functions with argument $(-i\Omega )$ give poles in the $\Omega $ integral only below the real axis.

For late times $u \to \infty $ we close the contour below and get at leading order
\begin{equation}
\begin{split}
-\pi i & \int d\Omega _1d\Omega _2 \, 
\exp \Big[ \big(2\pi -4\pi iT \td_{31} \big)\Omega  _1 +\big(2\pi-4\pi iT \td_{42} \big)\Omega _2 \Big] \\
& \Gamma (\Delta _1\pm i\Omega _1)\Gamma (\Delta _2 \pm i\Omega _2 ) \Gamma (1-\Delta _1\pm i\Omega _1  )\Gamma (1-\Delta _2 \pm i\Omega _2 ) \\
& \Bigg[ 4\pi iTt+4\pi iT \td_{43} -\pi+2i\gamma-2i\log(2s_0)  \\
& + i \frac{\Gamma '(\Delta _1+i\Omega _1)}{\Gamma (\Delta _1+i\Omega _1)} + i \frac{\Gamma '(\Delta _2-i\Omega _2)}{\Gamma (\Delta _2-i\Omega _2)} 
+ i \frac{\Gamma '(1-\Delta _1+i\Omega _1)}{\Gamma (1-\Delta _1+i\Omega _1)} + i \frac{\Gamma '(1-\Delta _2-i\Omega _2)}{\Gamma (1-\Delta _2-i\Omega _2)}  \Bigg]
\end{split}
\end{equation}
where $\gamma $ is Euler's constant.

For $u \to -\infty $ we should pick up the poles at $\Omega =in$ with $n=0,1,2,\cdots $. The leading contribution is from $n=0$ which is just minus the $u \to \infty $ result. The next subleading term comes from $n=1$ and gives
\begin{equation} \label{eq:crossed_first_term_n_1_pole}
\begin{split}
& -\frac{2\pi i}{4s_0^2} e^{4\pi Tt+4\pi T\td_{43} } \int d\Omega _1d\Omega _2 \, 
\exp \Big[ \big(2\pi -4\pi iT \td_{31} \big)\Omega  _1 +\big(2\pi-4\pi iT \td_{42} \big)\Omega _2 \Big] \\
& \Gamma (\Delta _1\pm i\Omega _1)\Gamma (\Delta _2 \pm i\Omega _2 ) \Gamma (1-\Delta _1\pm i\Omega _1  )\Gamma (1-\Delta _2 \pm i\Omega _2 ) \\
& (\Delta _1+i\Omega _1)(\Delta _2-i\Omega _2)(1-\Delta _1+i\Omega _1)(1-\Delta _2-i\Omega _2)
\\
& \Bigg[ 4\pi iTt+4\pi iT \td_{43} -\pi+2i(\gamma-1) -2i\log(2s_0) \\
& + i \frac{\Gamma '(\Delta _1+i\Omega _1+1)}{\Gamma (\Delta _1+i\Omega _1+1)} + i \frac{\Gamma '(\Delta _2-i\Omega _2+1)}{\Gamma (\Delta _2-i\Omega _2+1)} 
+ i \frac{\Gamma '(2-\Delta _1+i\Omega _1)}{\Gamma (2-\Delta _1+i\Omega _1)} + i \frac{\Gamma '(2-\Delta _2-i\Omega _2)}{\Gamma (2-\Delta _2-i\Omega _2)}  \Bigg] .
\end{split}
\end{equation}

\textbf{Second term} (first mixed term). This term is given by
\begin{equation}
\begin{split}
& \int \prod _{i=1} ^3 \left[ d\omega _i  \right]
\exp \Big[ (\pi -4\pi iT t_{21} )\omega _1 +(\pi -4\pi iT t_{32} )\omega _2 +(\pi -4\pi iT t_{43} )\omega _3 \Big] \\
& \Gamma (\Delta _1+i\omega _{23} )\Gamma (\Delta _1-i\omega _1)\Gamma (\Delta _2+i\omega _{21} )\Gamma (\Delta _2-i\omega _3)\Gamma \left( i(\omega _1+\omega _3-\omega _2)\right) \\
& \Gamma (1-\Delta _1+i\omega _{32} )\Gamma (1-\Delta _1+i\omega _1)\Gamma (1-\Delta _2+i\omega _{12} )\Gamma (1-\Delta _2+i\omega _3)\Gamma \left( i(\omega _2-\omega _1-\omega _3)\right) .
\end{split}
\end{equation}
With the same change of variables we have
\begin{equation}
\begin{split}
& \int d\Omega d\Omega _1d\Omega _2 \\
& \exp \Big[ \big(\pi-2i\log(2s_0)-4\pi iTu-4\pi iT \td_{43} \big)\Omega  +\big(2\pi -4\pi iT \td_{31} \big)\Omega  _1 +\big(2\pi-4\pi iT \td_{42} \big)\Omega _2 \Big] \\
& \Gamma (\Delta _1+i\Omega _1-i\Omega  )\Gamma (\Delta _1-i\Omega _1)\Gamma (\Delta _2+i\Omega _2 )\Gamma (\Delta _2-i\Omega _2-i\Omega )\Gamma (i\Omega )\Gamma(-i\Omega ) \\
& \Gamma (1-\Delta _1-i\Omega _1+i\Omega  )\Gamma (1-\Delta _1+i\Omega _1)\Gamma (1-\Delta _2-i\Omega _2 )\Gamma (1-\Delta _2+i\Omega _2+i\Omega ).
\end{split}
\end{equation}
The $s_0$ dependence means that the $\Omega $ integral concentrates around $\Omega =0$.
In addition to \eqref{eq:lim_of_cos_sin}, we have
\begin{equation} \label{eq:lim_of_cos_sin_2}
\frac{\cos(M\Omega )}{\Omega ^2}  \xrightarrow{M \to \infty } -M\pi \delta (\Omega ),\qquad \frac{\sin(M\Omega )}{\Omega^2 } \xrightarrow{M \to \infty } -\pi \delta' (\Omega ).
\end{equation}
This can be checked either directly or by taking a derivative of \eqref{eq:lim_of_cos_sin}. Therefore
\begin{equation}
\Gamma (i\Omega )\Gamma (-i\Omega ) e^{-2i\log(2s_0)\Omega } \sim -2\pi \log(2s_0)\delta (\Omega )+i\pi  \delta '(\Omega ).
\end{equation}
The second term then becomes
\begin{equation}
\begin{split}
\pi i & \int d\Omega _1d\Omega _2 \, 
\exp \Big[ \big(2\pi -4\pi iT \td_{31} \big)\Omega  _1 +\big(2\pi-4\pi iT \td_{42} \big)\Omega _2 \Big] \\
& \Gamma (\Delta _1\pm i\Omega _1)\Gamma (\Delta _2 \pm i\Omega _2 ) \Gamma (1-\Delta _1\pm i\Omega _1  )\Gamma (1-\Delta _2 \pm i\Omega _2 ) \\
& \Bigg[ 4\pi iTt+4\pi iT \td_{43} -\pi  \\
& + i \frac{\Gamma '(\Delta _1+i\Omega _1)}{\Gamma (\Delta _1+i\Omega _1)} + i \frac{\Gamma '(\Delta _2-i\Omega _2)}{\Gamma (\Delta _2-i\Omega _2)} 
- i \frac{\Gamma '(1-\Delta _1-i\Omega _1)}{\Gamma (1-\Delta _1-i\Omega _1)} - i \frac{\Gamma '(1-\Delta _2+i\Omega _2)}{\Gamma (1-\Delta _2+i\Omega _2)}  \Bigg] .
\end{split}
\end{equation}

\textbf{Third term} (second mixed term). This is just $\Delta _i \to 1-\Delta _i$ of the previous term.

\textbf{Fourth term}. 
This term explicitly is
\begin{equation}
\begin{split}
& \int \prod _{i=1} ^3 \left[ d\omega _i  \right]
\exp \Big[ (\pi -2i\log(2s_0)-4\pi iT t_{21} )\omega _1 +(\pi +2i\log(2s_0)-4\pi iT t_{32} )\omega _2 \\
& \qquad \qquad +(\pi -2i\log(2s_0)-4\pi iT t_{43} )\omega _3 \Big] \\
& \Gamma (\Delta _1+i\omega _{32} )\Gamma (\Delta _1+i\omega _1)\Gamma (\Delta _2+i\omega _{12} )\Gamma (\Delta _2+i\omega _3)\Gamma \left( i(\omega _2-\omega _1-\omega _3)\right)^2 \\
& \Gamma (1-\Delta _1+i\omega _{32} )\Gamma (1-\Delta _1+i\omega _1)\Gamma (1-\Delta _2+i\omega _{12} )\Gamma (1-\Delta _2+i\omega _3) .
\end{split}
\end{equation}
The same change of variables leads to
\begin{equation}
\begin{split}
& \int d\Omega d\Omega _1d\Omega _2 \\
& \exp \Big[ \big(\pi-4i\log(2s_0)-4\pi iTu-4\pi iT \td_{43} \big)\Omega  +\big(2\pi -4\pi iT \td_{31} \big)\Omega  _1 +\big(2\pi-4\pi iT \td_{42} \big)\Omega _2 \Big] \\
& \Gamma (\Delta _1-i\Omega _1+i\Omega  )\Gamma (\Delta _1+i\Omega _1)\Gamma (\Delta _2-i\Omega _2 )\Gamma (\Delta _2+i\Omega _2+i\Omega )\Gamma (-i\Omega )^2 \\
& \Gamma (1-\Delta _1-i\Omega _1+i\Omega  )\Gamma (1-\Delta _1+i\Omega _1)\Gamma (1-\Delta _2-i\Omega _2 )\Gamma (1-\Delta _2+i\Omega _2+i\Omega ).
\end{split}
\end{equation}
Once again we localize around $\Omega =0$ and use that
\begin{equation}
\Gamma (-i\Omega )^2 e^{-4i\log(2s_0)\Omega } \sim 4\pi \log(2s_0)\delta (\Omega )-i\pi \delta '(\Omega )-2\pi \gamma \delta (\Omega )
\end{equation}
to obtain
\begin{equation}
\begin{split}
\pi i & \int d\Omega _1d\Omega _2 \, 
\exp \Big[ \big(2\pi -4\pi iT \td_{31} \big)\Omega  _1 +\big(2\pi-4\pi iT \td_{42} \big)\Omega _2 \Big] \\
& \Gamma (\Delta _1\pm i\Omega _1)\Gamma (\Delta _2 \pm i\Omega _2 ) \Gamma (1-\Delta _1\pm i\Omega _1  )\Gamma (1-\Delta _2 \pm i\Omega _2 ) \\
& \Bigg[ -4\pi iTt-4\pi iT \td_{43} +\pi-2i\log(2s_0)+2\gamma i \\
& + i \frac{\Gamma '(\Delta _1-i\Omega _1)}{\Gamma (\Delta _1-i\Omega _1)} + i \frac{\Gamma '(\Delta _2+i\Omega _2)}{\Gamma (\Delta _2+i\Omega _2)} 
+ i \frac{\Gamma '(1-\Delta _1-i\Omega _1)}{\Gamma (1-\Delta _1-i\Omega _1)} + i \frac{\Gamma '(1-\Delta _2+i\Omega _2)}{\Gamma (1-\Delta _2+i\Omega _2)}  \Bigg] .
\end{split}
\end{equation}

\textbf{Putting it all together}. For very long times all terms cancel. This is the expected behavior of an out-of-time-ordered correlator. It is not obvious in our setup since the operators are time-integrated operators and one could worry that we lose this property. We see that there is no such problem.

For times earlier than the scrambling time, we can organize the result as follows. At leading order, we have the $n=0$ pole contribution from the first term plus the full result of all the other three terms. These give together
\begin{equation} \label{eq:crossed_4pf_leading_behavior}
\begin{split}
2\pi i & \int d\Omega _1d\Omega _2 \, 
\exp \Big[ \big(2\pi -4\pi iT \td_{31} \big)\Omega  _1 +\big(2\pi-4\pi iT \td_{42} \big)\Omega _2 \Big] \\
& \Gamma (\Delta _1\pm i\Omega _1)\Gamma (\Delta _2 \pm i\Omega _2 ) \Gamma (1-\Delta _1\pm i\Omega _1  )\Gamma (1-\Delta _2 \pm i\Omega _2 ) \\
& \Bigg[ 4\pi iTt+4\pi iT \td_{43} -\pi+2i\gamma-2i\log(2s_0)  \\
& + i \frac{\Gamma '(\Delta _1+i\Omega _1)}{\Gamma (\Delta _1+i\Omega _1)} + i \frac{\Gamma '(\Delta _2-i\Omega _2)}{\Gamma (\Delta _2-i\Omega _2)} 
+ i \frac{\Gamma '(1-\Delta _1+i\Omega _1)}{\Gamma (1-\Delta _1+i\Omega _1)} + i \frac{\Gamma '(1-\Delta _2-i\Omega _2)}{\Gamma (1-\Delta _2-i\Omega _2)}  \Bigg]
\end{split}
\end{equation}

All subleading terms come from the first term. At the first subleading order we have just the $n=1$ pole from the first term which we wrote before in \eqref{eq:crossed_first_term_n_1_pole}.

Now, in our holographic-semiclassical limit, combined with early times compared to the scrambling time, the $\log(2s_0)$ term is most dominant, and the $t$-dependent term cannot be comparable to it, which would contradict the early times assumption. The fact that we have a linear in $t$ dependence (and in particular not only the expansion in exponentials of $t$) is non-standard, but we see that this dependence is subleading. We find a surprising $\log(s_0)$ dependence in our case. Explicitly, the leading behavior of the early time regime, restoring the coefficient in the first line of \eqref{eq:crossed_4pf_int_expr_before_simplify} is
\begin{equation} \label{eq:crossed_4pf_int_expr}
\begin{split}
& \langle \cO _2(t_4) \cO _1(t_3) \cO _2(t_2) \cO _1(t_1) \rangle = \frac{(2\lambda s_0)^{2\Delta _1+2(1-\Delta _1)+2\Delta _2+2(1-\Delta _2)} }
{\Gamma (2\Delta _1) \Gamma (2\Delta _2)\Gamma (2-2\Delta _1)\Gamma (2-2\Delta _2) \big(16 \pi ^3T\big)^{3}} 4 \pi \log(s_0)\\
& \bigg\{
\int d\Omega _1d\Omega _2 \, 
\exp \Big[ \big(2\pi -4\pi iT \td_{31} \big)\Omega  _1 +\big(2\pi-4\pi iT \td_{42} \big)\Omega _2 \Big] \\
& \qquad \Gamma (\Delta _1\pm i\Omega _1)\Gamma (\Delta _2 \pm i\Omega _2 ) \Gamma (1-\Delta _1\pm i\Omega _1  )\Gamma (1-\Delta _2 \pm i\Omega _2 ) \\
& -\frac{1}{4s_0^2} e^{4\pi Tt+4\pi T\td_{43} } \int d\Omega _1d\Omega _2 \, 
\exp \Big[ \big(2\pi -4\pi iT \td_{31} \big)\Omega  _1 +\big(2\pi-4\pi iT \td_{42} \big)\Omega _2 \Big] \\
& \qquad \Gamma (\Delta _1\pm i\Omega _1)\Gamma (\Delta _2 \pm i\Omega _2 ) \Gamma (1-\Delta _1\pm i\Omega _1  )\Gamma (1-\Delta _2 \pm i\Omega _2 ) \\
& \qquad (\Delta _1+i\Omega _1)(\Delta _2-i\Omega _2)(1-\Delta _1+i\Omega _1)(1-\Delta _2-i\Omega _2) + \cdots \bigg\} .
\end{split}
\end{equation}
In the main text we write the result in terms of the original non-redundant time variables, which is simply obtained by taking $t \to 0$ and $\td_i \to t_i$.

\textbf{Simplifying the expression}.
We see that we essentially have three kinds of integrals. The first integral is the one appearing in the 2-point function
\begin{equation}
I_{\Delta } ^{(1)} (t ) \equiv \int d\Omega \, e^{(2\pi -2it)\Omega } \Gamma (\Delta \pm i\Omega )\Gamma (1-\Delta \pm i\Omega ) \propto \frac{\sinh\left( \mu (t +i\pi )\right) }{\sinh(t )} \qquad \text{with }\, \mu =1-2\Delta 
\end{equation}
or, more precise, as we have seen
\begin{equation}
\langle \cO (t) \cO (0)\rangle =\frac{(2\lambda s_0)^{2\Delta +2(1-\Delta )} }{16\pi ^3 T\, \Gamma (2\Delta )\Gamma (2-2\Delta )} I^{(1)} _{\Delta } (2\pi T t) .
\end{equation}
We also have
\begin{equation}
I_{\Delta } ^{(2)} (t) \equiv \int d\Omega \, e^{(2\pi -2it)\Omega } \Gamma (\Delta \pm i\Omega )\Gamma (1-\Delta  \pm i\Omega )(\Delta +i\Omega )(1-\Delta +i\Omega )
\end{equation}
and
\begin{equation}
I_{\Delta } ^{(3)} (t) \equiv \int d\Omega \, e^{(2\pi -2it)\Omega } \Gamma (\Delta \pm i\Omega )\Gamma (1-\Delta  \pm i\Omega )(\Delta -i\Omega )(1-\Delta -i\Omega ) .
\end{equation}
The ratios between these integrals is
\begin{equation}
\begin{split}
& \frac{I^{(2)} _{\Delta } (t)}{I^{(1)} _{\Delta } (t)} =\\
& = \frac{1}{2} +\Big( \Delta -\frac{1}{2} \Big) \coth \left( \mu (t+i\pi )\right) +\frac{1}{2} \coth t+\Big( \Delta -\frac{1}{2} \Big) \coth \left( \mu (t+i\pi )\right)  \coth (t) +\frac{1}{2} \left( \sinh(t)\right) ^{-2} \\
&=\frac{1}{2} \left( \coth t+1\right) \Big[ \coth t + (2\Delta -1) \coth \left( \mu (t+i\pi )\right) \Big] 
\end{split}
\end{equation}
and
\begin{equation}
\begin{split}
&\frac{I^{(3)} _{\Delta } (t)}{I^{(1)} _{\Delta } (t)} =\\
&=\frac{1}{2} -\Big( \Delta -\frac{1}{2} \Big) \coth \left( \mu (t+i\pi )\right) -\frac{1}{2} \coth t+\Big( \Delta -\frac{1}{2} \Big) \coth \left( \mu (t+i\pi )\right)  \coth (t) +\frac{1}{2} \left( \sinh(t)\right) ^{-2}\\
&=\frac{1}{2} \left( \coth t-1\right) \Big[ \coth t + (2\Delta -1) \coth \left( \mu (t+i\pi )\right) \Big] .
\end{split}
\end{equation}
In the main text we denote these two ratios by $\ratiocoeff_{\Delta } (t)$ and $\ratiocoeff'_{\Delta } (t)$ respectively.

In terms of these rather simple functions, we can write \eqref{eq:crossed_4pf_int_expr}, or better the normalized 4-point function, as
\begin{equation}
\begin{split}
& \frac{\langle \cO _2(t_4) \cO _1(t_3) \cO _2(t_2) \cO _1(t_1) \rangle}{\langle \cO _2(t_4) \cO _2(t_2)\rangle \langle \cO _1(t_3) \cO _1(t_1)\rangle } = 
\frac{\log s_0}{4\pi ^2T} \bigg\{ 1
-\frac{e^{4\pi T(t+ \td_{43}) } }{4s_0^2} \frac{I^{(2)} _{\Delta _1} }{I^{(1)} _{\Delta _1} } (2\pi T \td_{31} ) \cdot \frac{I^{(3)} _{\Delta _2} }{I^{(1)} _{\Delta _2} } (2\pi T\td_{42} ) + \cdots \bigg\} .
\end{split}
\end{equation}

\subsection{Finite times} \label{sec:finite_times}

For completeness, let us consider what happens for finite times, not close to the scrambling time. We will similarly analyze the four terms as before. In this case, we do not have the time regime \eqref{eq:crossed_scrambling_time_cov}, and rather keep generic finite $t_i$. However, it will still be useful to do the change of variables \eqref{eq:crossed_4pf_pert_evaluation_energies_cov} for the frequencies which isolates the oscillatory behavior.

For the first term, we start with \eqref{eq:crossed_4pf_first_term} and with the change of variables \eqref{eq:crossed_4pf_pert_evaluation_energies_cov} we get, instead of \eqref{eq:crossed_4pf_first_term_after_cov}, the following
\begin{equation}
\begin{split}
& \int d\Omega d\Omega _1d\Omega _2 
 \exp \Big[ \big(\pi+2i\log(2s_0)-4\pi iT t_{43} \big)\Omega  +\big(2\pi -4\pi iT t_{31} \big)\Omega  _1 +\big(2\pi-4\pi iT t_{42} \big)\Omega _2 \Big] \\
& \Gamma (\Delta _1+i\Omega _1-i\Omega  )\Gamma (\Delta _1-i\Omega _1)\Gamma (\Delta _2+i\Omega _2 )\Gamma (\Delta _2-i\Omega _2-i\Omega )\Gamma (i\Omega )^2 \\
& \Gamma (1-\Delta _1+i\Omega _1-i\Omega  )\Gamma (1-\Delta _1-i\Omega _1)\Gamma (1-\Delta _2+i\Omega _2 )\Gamma (1-\Delta _2-i\Omega _2-i\Omega ).
\end{split}
\end{equation}
This localizes to $\Omega =0$ and using
\begin{equation}
\Gamma (i\Omega )^2 e^{2i\log(2s_0)\Omega } \sim 2\pi \log(2s_0)\delta (\Omega )+i\pi \delta '(\Omega )-2\pi \gamma \delta (\Omega )
\end{equation}
we get
\begin{equation}
\begin{split}
\pi i & \int d\Omega _1d\Omega _2 \, 
\exp \Big[ \big(2\pi -4\pi iT t_{31} \big)\Omega  _1 +\big(2\pi-4\pi iT t_{42} \big)\Omega _2 \Big] \\
& \Gamma (\Delta _1\pm i\Omega _1)\Gamma (\Delta _2 \pm i\Omega _2 ) \Gamma (1-\Delta _1\pm i\Omega _1  )\Gamma (1-\Delta _2 \pm i\Omega _2 ) 
 \Bigg[ 4\pi iT t_{43} -\pi+2i\gamma-2i\log(2s_0)  \\
& + i \frac{\Gamma '(\Delta _1+i\Omega _1)}{\Gamma (\Delta _1+i\Omega _1)} + i \frac{\Gamma '(\Delta _2-i\Omega _2)}{\Gamma (\Delta _2-i\Omega _2)} 
+ i \frac{\Gamma '(1-\Delta _1+i\Omega _1)}{\Gamma (1-\Delta _1+i\Omega _1)} + i \frac{\Gamma '(1-\Delta _2-i\Omega _2)}{\Gamma (1-\Delta _2-i\Omega _2)}  \Bigg].
\end{split}
\end{equation}

For the second term we get instead
\begin{equation}
\begin{split}
& \int d\Omega d\Omega _1d\Omega _2 
 \exp \Big[ \big(\pi-4\pi iT t_{43} \big)\Omega  +\big(2\pi -4\pi iT t_{31} \big)\Omega  _1 +\big(2\pi-4\pi iT t_{42} \big)\Omega _2 \Big] \\
& \Gamma (\Delta _1+i\Omega _1-i\Omega  )\Gamma (\Delta _1-i\Omega _1)\Gamma (\Delta _2+i\Omega _2 )\Gamma (\Delta _2-i\Omega _2-i\Omega )\Gamma (i\Omega ) \Gamma (-i\Omega ) \\
& \Gamma (1-\Delta _1-i\Omega _1+i\Omega  )\Gamma (1-\Delta _1+i\Omega _1)\Gamma (1-\Delta _2-i\Omega _2 )\Gamma (1-\Delta _2+i\Omega _2+i\Omega ).
\end{split}
\end{equation}
and the third term is $\Delta _i \to 1-\Delta _i$ of this. Curiously these two terms do not simplify. In particular they are not given by a factorized answer, which is what we would have expected. We will come back to this.

In the fourth term we have
\begin{equation}
\begin{split}
& \int d\Omega d\Omega _1d\Omega _2 
 \exp \Big[ \big(\pi-2i\log(2s_0)-4\pi iT t_{43} \big)\Omega  +\big(2\pi -4\pi iT t_{31} \big)\Omega  _1 +\big(2\pi-4\pi iT t_{42} \big)\Omega _2 \Big] \\
& \Gamma (\Delta _1-i\Omega _1+i\Omega  )\Gamma (\Delta _1+i\Omega _1)\Gamma (\Delta _2-i\Omega _2 )\Gamma (\Delta _2+i\Omega _2+i\Omega )\Gamma (-i\Omega )^2 \\
& \Gamma (1-\Delta _1-i\Omega _1+i\Omega  )\Gamma (1-\Delta _1+i\Omega _1)\Gamma (1-\Delta _2-i\Omega _2 )\Gamma (1-\Delta _2+i\Omega _2+i\Omega ).
\end{split}
\end{equation}
Using
\begin{equation}
\Gamma (-i\Omega )^2 e^{-2i\log(2s_0)\Omega } \sim 2\pi \log(2s_0)\delta (\Omega )-i\pi \delta '(\Omega )-2\pi \gamma \delta (\Omega )
\end{equation}
the fourth term simplifies to
\begin{equation}
\begin{split}
\pi i & \int d\Omega _1d\Omega _2 \, 
\exp \Big[ \big(2\pi -4\pi iT t_{31} \big)\Omega  _1 +\big(2\pi-4\pi iT t_{42} \big)\Omega _2 \Big] \\
& \Gamma (\Delta _1\pm i\Omega _1)\Gamma (\Delta _2 \pm i\Omega _2 ) \Gamma (1-\Delta _1\pm i\Omega _1  )\Gamma (1-\Delta _2 \pm i\Omega _2 )
 \Bigg[ -4\pi iT t_{43} +\pi-2i\log(2s_0)+2\gamma i \\
& + i \frac{\Gamma '(\Delta _1-i\Omega _1)}{\Gamma (\Delta _1-i\Omega _1)} + i \frac{\Gamma '(\Delta _2+i\Omega _2)}{\Gamma (\Delta _2+i\Omega _2)} 
+ i \frac{\Gamma '(1-\Delta _1-i\Omega _1)}{\Gamma (1-\Delta _1-i\Omega _1)} + i \frac{\Gamma '(1-\Delta _2+i\Omega _2)}{\Gamma (1-\Delta _2+i\Omega _2)}  \Bigg] .
\end{split}
\end{equation}

While the second and third terms do not simplify, we note that in the holographic-semiclassical limit they are in fact negligible compared to the other terms, and the leading behavior of the sum of all terms is
\begin{equation}
\begin{split}
\pi i \big(-4i \log(2s_0) \big) & \int d\Omega _1d\Omega _2 \, 
\exp \Big[ \big(2\pi -4\pi iT t_{31} \big)\Omega  _1 +\big(2\pi-4\pi iT t_{42} \big)\Omega _2 \Big] \\
& \Gamma (\Delta _1\pm i\Omega _1)\Gamma (\Delta _2 \pm i\Omega _2 ) \Gamma (1-\Delta _1\pm i\Omega _1  )\Gamma (1-\Delta _2 \pm i\Omega _2 ) .
\end{split}
\end{equation}
We expect this result to agree with the very early time behavior in the scrambling regime. Indeed, we see that this agrees with the dominant behavior of \eqref{eq:crossed_4pf_leading_behavior}. 

Including the coefficient in the first line of \eqref{eq:crossed_4pf_int_expr_before_simplify}, this result agrees with $\langle \cO _2(t_4) \cO _2(t_2)\rangle \langle \cO _1(t_3) \cO _1(t_1)\rangle $, up to an additional factor of $ \frac{\log s_0}{4\pi ^2T} $. We see that in terms of the time dependence, we get the desired factorization into a product of 2-point functions in the semiclassical limit. However, this happens in a rather non-trivial way.

\subsection{Perturbative evaluation for more general operators} \label{sec:crossed_pert_evaluation_general_d}

We repeat the perturbative analysis of Sec.\ \ref{sec:crossed_pert_evaluation} where now each of the four operators is of the form \eqref{eq:doubled_generic_ops} with parameters \eqref{eq:dimensions_giving_dS_d} corresponding to de Sitter space of arbitrary dimension $d$. We denote $\delta =\frac{d-1}{2} $ and so parameterize the first operator $\cO _1$ by $\Delta ^{(1)} \to \Delta _1$ and $\Delta ^{(2)} \to \delta -\Delta _1$ in \eqref{eq:doubled_generic_ops}, and similarly for $\cO _2$ we take $\Delta _2$ and $\delta -\Delta _2$.

The starting point, eq.\ \eqref{eq:crossed_4pf_int_expr_before_simplify}, becomes for such operators
\begin{equation}
\begin{split}
& \langle \cO _2(t_4) \cO _1(t_3) \cO _2(t_2) \cO _1(t_1) \rangle = \\
& = (2\lambda s_0)^{2\Delta _1+2(\delta -\Delta _1)+2\Delta _2+2(\delta -\Delta _2)}
\Big[ \Gamma (2\Delta _1) \Gamma (2\Delta _2)\Gamma (2\delta -2\Delta _1)\Gamma (2\delta -2\Delta _2)\Big]^{-1} \big(16 \pi ^3T\big)^{-3} \\
& \int \prod _{i=1} ^3 \left[ d\omega _i  \right]
\exp \Big[ (\pi -4\pi iT t_{21} )\omega _1 +(\pi -4\pi iT t_{32} )\omega _2 +(\pi -4\pi iT t_{43} )\omega _3 \Big] \\
\bigg\{
& \Gamma (\Delta _1+i\omega _{23} )\Gamma (\Delta _1-i\omega _1)\Gamma (\Delta _2+i\omega _{21} )\Gamma (\Delta _2-i\omega _3)\Gamma \left( i(\omega _1+\omega _3-\omega _2)\right) (2s_0)^{i(\omega _1-\omega _2+\omega _3)} \\
& + \Gamma (\Delta _1+i\omega _{32} )\Gamma (\Delta _1+i\omega _1)\Gamma (\Delta _2+i\omega _{12} )\Gamma (\Delta _2+i\omega _3)\Gamma \left( i(\omega _2-\omega _1-\omega _3)\right) (2s_0)^{i(-\omega _1+\omega _2-\omega _3)} \bigg\} \\
& \cdot (\Delta _i \to \delta -\Delta _i) .
\end{split}
\end{equation}

At very long times (which we obtained by $u \to \infty $), the OTOC still goes to zero. At times much before the scrambling time ($u \to - \infty $), we get the following. At leading order we have the $\Omega =0$ pole from the first term, plus the second, third and fourth terms discussed in Sec.\ \ref{sec:crossed_pert_evaluation}; this sum corresponds to $n=0$ in the following formula. Then we have all the $\Omega =in$ poles, with $n=1,2, \cdots $, coming from the first term. We get
\begin{equation}
\begin{split}
& \langle \cO _2(t_4) \cO _1(t_3) \cO _2(t_2) \cO _1(t_1) \rangle = \frac{(2\lambda s_0)^{2\Delta _1+2(\delta -\Delta _1)+2\Delta _2+2(\delta -\Delta _2)} }
{\Gamma (2\Delta _1) \Gamma (2\Delta _2)\Gamma (2\delta -2\Delta _1)\Gamma (2\delta -2\Delta _2) \big(16 \pi ^3T\big)^{3}} \\
& \cdot 2\pi i \sum _{n=0} ^{\infty } \frac{(-1)^n}{(n!)^2} \frac{1}{(2s_0)^{2n} } e^{4\pi n T(t+\td_{43}) } \\
&
\int d\Omega _1d\Omega _2 \, 
\exp \Big[ \big(2\pi -4\pi iT \td_{31} \big)\Omega  _1 +\big(2\pi-4\pi iT \td_{42} \big)\Omega _2 \Big] \\
& \qquad \Gamma (\Delta _1+ i\Omega _1+n)\Gamma (\Delta _1-i\Omega _1)\Gamma (\Delta _2 + i\Omega _2 )\Gamma (\Delta _2-i\Omega _2+n) \\
& \qquad \Gamma (\delta -\Delta _1+ i\Omega _1+n)\Gamma (\delta -\Delta _1-i\Omega _1)\Gamma (\delta -\Delta _2 + i\Omega _2 )\Gamma (\delta -\Delta _2-i\Omega _2+n) \\
& \Bigg[ -2i \frac{\Gamma '(n+1)}{\Gamma (n+1)} -\pi +4\pi iTt-2i\log(2s_0)+4\pi iT\td_{43} \\
& +i \frac{\Gamma '(\Delta _1+i\Omega _1+n)}{\Gamma (\Delta _1+i\Omega _1+n)} +i \frac{\Gamma '(\Delta _2-i\Omega _2+n)}{\Gamma (\Delta _2-i\Omega _2+n)} 
+i \frac{\Gamma '(\delta -\Delta _1+i\Omega _1+n)}{\Gamma (\delta -\Delta _1+i\Omega _1+n)} +i \frac{\Gamma '(\delta -\Delta _2-i\Omega _2+n)}{\Gamma (\delta -\Delta _2-i\Omega _2+n)} \Bigg].
\end{split}
\end{equation}
Note that $\frac{\Gamma '(n+1)}{\Gamma (n+1)} = - \gamma +\sum _{k=1} ^n \frac{1}{k} $ (and for $n=0$ it is just $-\gamma $). Keeping only the dominant $s_0$ term this becomes
\begin{equation}
\begin{split}
& \langle \cO _2(t_4) \cO _1(t_3) \cO _2(t_2) \cO _1(t_1) \rangle = \frac{(2\lambda s_0)^{2\Delta _1+2(\delta -\Delta _1)+2\Delta _2+2(\delta -\Delta _2)} }
{\Gamma (2\Delta _1) \Gamma (2\Delta _2)\Gamma (2\delta -2\Delta _1)\Gamma (2\delta -2\Delta _2) \big(16 \pi ^3T\big)^{3}} \, 4\pi \log(s_0) \\
& \cdot \sum _{n=0} ^{\infty } \frac{(-1)^n}{(n!)^2} \frac{1}{(2s_0)^{2n} } e^{4\pi n T(t+\td_{43}) } \\
&
\int d\Omega _1d\Omega _2 \, 
\exp \Big[ \big(2\pi -4\pi iT \td_{31} \big)\Omega  _1 +\big(2\pi-4\pi iT \td_{42} \big)\Omega _2 \Big] \\
& \qquad \Gamma (\Delta _1+ i\Omega _1+n)\Gamma (\Delta _1-i\Omega _1)\Gamma (\Delta _2 + i\Omega _2 )\Gamma (\Delta _2-i\Omega _2+n) \\
& \qquad \Gamma (\delta -\Delta _1+ i\Omega _1+n)\Gamma (\delta -\Delta _1-i\Omega _1)\Gamma (\delta -\Delta _2 + i\Omega _2 )\Gamma (\delta -\Delta _2-i\Omega _2+n) .
\end{split}
\end{equation}

If we change variables $\Omega _1=\Omega '_1+\frac{in}{2} $ and $\Omega _2=\Omega '_2-\frac{in}{2} $ we get the integrals we encountered for the 2-point function. That is, after this change of variables, we get for each of the two flavors of operators an integral of the form \eqref{eq:Fourier_integral_giving_2F1}, which we can then relate to the 2-point function in de Sitter using \eqref{eq:2pf_dS_2F1}. Given that $\cO _1$ has parameters $\Delta _1$ and $\delta -\Delta _1$ related to a particle with $\mud_1$ and a bulk of dimension $d$ through \eqref{eq:dimensions_giving_dS_d}, the change of variables gives us instead the parameters $\Delta _1+\frac{n}{2} $ and $\delta -\Delta _1+\frac{n}{2} $, which we can think of as a particle with the same $\mud_1$ but in dimension $d+2n$. The same applies to $\cO _2$.  So we can express the result using the 2-point functions in de Sitter space in various bulk dimensions. The normalized 4-point function written in terms of the non-redundant time variables is then
\begin{equation}
\begin{split}
& \frac{\langle \cO _2(t_4) \cO _1(t_3) \cO _2(t_2) \cO _1(t_1) \rangle}{\langle \cO _2(t_4) \cO _2(t_2)\rangle \langle \cO _1(t_3) \cO _1(t_1)\rangle }= \frac{\log s_0}{4\pi ^2 \, T} 
\sum _{n=0} ^{\infty } \frac{(-1)^n}{(n!)^2} \frac{1}{(2s_0)^{2n}} e^{2\pi nT(2t_{43} +t_{31} -t_{42} )} \\
& \qquad \qquad \frac{\Gamma \left( \frac{d+2n-1}{2} \right) ^4 \Gamma \left( \frac{d}{2} +n\right) ^2}{\pi ^{2n} T^{4n}\, \Gamma (d+2n-1)^2} \frac{\Gamma (d-1)^2}{\Gamma \left( \frac{d-1}{2} \right) ^4 \Gamma \left( \frac{d}{2} \right) ^2}  \frac{G_{d+2n,\mud _1} (t_{31} )G_{d+2n,\mud _2} (t_{42} )}{G_{d,\mud _1} (t_{31} )G_{d,\mud _2} (t_{42} )} .
\end{split}
\end{equation}

\section{Limit of a very-well-poised basic hypergeometric series} \label{sec:limit_of_W}

We again use a notation where $s_{\pm i \pm j \pm \cdots } =\pm s_i \pm s_j \pm \cdots $.
For generic values of the parameters we can use the Mellin-Barnes-Agarwal integral representation of the ${}_8W_7$ function
\begin{equation}
\begin{split}
& {}_8W_7 \big( q^{\Delta _1+2\Delta _2-1+is_{-2-3} } ;q^{\Delta _2+is_{1-2} } ,q^{\Delta _2+is_{-3 - 4} },q^{\Delta _2+is_{4-3} }  ,q^{\Delta _2+is_{-1-2} } ,q^{\Delta _1+is_{-2-3} } ;q,q^{\Delta _1+is_{2+3} } \big)=\\
& = -\Gamma  (\Delta _2-\Delta _1+1+is_{1-3} )^{-1}\Gamma  (\Delta _1-\Delta _2+is_{3-1} )^{-1} \\
& \frac{\Gamma _q(1+\Delta _2-\Delta _1+is_{1-3} )\Gamma _q(\Delta _1-\Delta _2+is_{3-1} )\Gamma _q(\Delta _1+\Delta _2+is_{-3\pm 1} )\Gamma _q(\Delta _1+\Delta _2+is_{-2 \pm 4} )\Gamma _q(2\Delta _2)}{\Gamma _q(\Delta _1+2\Delta _2+is_{-2-3} )\Gamma _q(\Delta _2+is_{1 \pm 2} )\Gamma _q(\Delta _2+is _{\pm 4-3} )\Gamma _q(\Delta _1+is_{-1 \pm 4} )\Gamma _q(\Delta _1+is_{3-2} )} \\
& \int \frac{d\uing}{2\pi i} q^{\uing} \Gamma (-\uing)\Gamma (1+\uing)\Gamma (1+\uing+\Delta _2-\Delta _1+is_{1-3} )\Gamma (-\uing+\Delta _1-\Delta _2+is_{3-1} ) \\
& \frac{\Gamma _q(\uing+\Delta _2+is_{1-2} )\Gamma _q(\uing+\Delta _2+is_{-3 \pm 4} )\Gamma _q(\uing+\Delta _2+is_{1+2} )}{\Gamma _q(\uing+1)\Gamma _q(\uing+\Delta _1+\Delta _2+is_{1-3} )\Gamma _q(\uing+2\Delta _2)\Gamma _q(\uing+1+\Delta _2-\Delta _1+is_{1-3} )} .
\end{split}
\end{equation}
The contour of integration is a deformation of a purely imaginary contour such that the poles coming from Gamma functions with arguments $(\uing+\cdots )$ are to the left of the contour and those from Gamma functions with $(-\uing+\cdots )$ are to the right.
We change variables to $\uing = \uint -\Delta _2+is_{3-4} $ and then
\begin{equation}
\begin{split}
& {}_8W_7 \big( q^{\Delta _1+2\Delta _2-1+is_{-2-3} } ;q^{\Delta _2+is_{1-2} } ,q^{\Delta _2+is_{-3 - 4} },q^{\Delta _2+is_{4-3} }  ,q^{\Delta _2+is_{-1-2} } ,q^{\Delta _1+is_{-2-3} } ;q,q^{\Delta _1+is_{2+3} } \big)=\\
& = -\Gamma  (\Delta _2-\Delta _1+1+is_{1-3} )^{-1}\Gamma  (\Delta _1-\Delta _2+is_{3-1} )^{-1} \\
& \frac{\Gamma _q(1+\Delta _2-\Delta _1+is_{1-3} )\Gamma _q(\Delta _1-\Delta _2+is_{3-1} )\Gamma _q(\Delta _1+\Delta _2+is_{-3\pm 1} )\Gamma _q(\Delta _1+\Delta _2+is_{-2 \pm 4} )\Gamma _q(2\Delta _2)}{\Gamma _q(\Delta _1+2\Delta _2+is_{-2-3} )\Gamma _q(\Delta _2+is_{1 \pm 2} )\Gamma _q(\Delta _2+is _{\pm 4-3} )\Gamma _q(\Delta _1+is_{-1 \pm 4} )\Gamma _q(\Delta _1+is_{3-2} )} \\
& \int \frac{d\uint}{2\pi i} q^{\uint -\Delta _2+is_{3-4} } \Gamma (-\uint+\Delta _2+is_{4-3} )\Gamma (\uint+1-\Delta _2+is_{3-4} ) \Gamma (\uint+1-\Delta _1+is_{1-4} )\Gamma (-\uint+\Delta _1+is_{4-1} ) \\
& \frac{\Gamma _q(\uint+is_{1+3-2-4} )\Gamma _q(\uint)\Gamma _q(\uint-2is_4)\Gamma _q(\uint+is_{1+2+3-4} )}{\Gamma _q(\uint+1-\Delta _2+is_{3-4} )\Gamma _q(\uint+\Delta _1+is_{1-4} )\Gamma _q(\uint+\Delta _2+is_{3-4} )\Gamma _(\uint+1-\Delta _1+is_{1-4} )} .
\end{split}
\end{equation}

Now we go to the holographic-semiclassical limit so that all $\Gamma _q$ are $\Gamma $ functions.
We use an argument similar to \cite{Lam:2018pvp}. In this case, the integral is of the form
\begin{equation}
\int d\uint \frac{\Gamma (\uint+A_1)\Gamma (\uint+A_2)\Gamma (\uint+A_3)\Gamma (\uint+A_4)\Gamma (-\uint+A_5)\Gamma (-\uint+A_6)}{\Gamma (\uint+B_1)\Gamma (\uint+B_2)} 
\end{equation}
where $A_1$ has a large positive imaginary part and $A_2$ has a large negative imaginary part of the same magnitude. We then notice that
\begin{equation}
\frac{|\Gamma (ix)|}{\sqrt{2\pi }} \sim \frac{1}{\sqrt{|x|}} e^{-\frac{\pi }{2} |x|} \qquad \text{for } x \in \mathbb{R} .
\end{equation}
So such Gamma functions are exponentially suppressed, and the exponential suppression is not affected by finite shifts in the arguments. For the pole at $\uint=-A_2$ we get in total $2+4-2=4$ exponential suppressions, where the first $2$ comes from $\Gamma (A_1-A_2)$ being doubly suppressed, and there are 4 additional ones in the numerator and 2 in the denominator. Similarly for $\uint=-A_1$. At poles other than $\uint=-A_1,-A_2$, such as at $\uint=-A_3$ we get only two exponential suppressions and so these are more dominant. At the other poles $\uint = -A_3-n$ with $n=1,2,3,\cdots $, we get further power law suppression compared to $n=0$ since for instance $\Gamma (A_1-A_3-1)= \frac{\Gamma (A_1-A_3)}{A_1-A_3-1} $. To summarize, if we close the contour to the left, we need to keep only the poles at $\uint=-A_3,-A_4$ which are the first poles with finite values in the corresponding series.

We conclude that in this limit, we have
\begin{equation} \label{eq:lim_of_8W7}
\begin{split}
& {}_8W_7 \big( q^{\Delta _1+2\Delta _2-1+is_{-2-3} } ;q^{\Delta _2+is_{1-2} } ,q^{\Delta _2+is_{-3 - 4} },q^{\Delta _2+is_{4-3} }  ,q^{\Delta _2+is_{-1-2} } ,q^{\Delta _1+is_{-2-3} } ;q,q^{\Delta _1+is_{2+3} } \big)\sim \\
& \sim \frac{\Gamma (\Delta _1+\Delta _2+is_{\pm 1-3} )\Gamma (\Delta _1+\Delta _2+is_{\pm 4-2} )\Gamma (2\Delta _2)\Gamma (is_{1+3-2-4} )\Gamma  (-2is_4)\Gamma (is_{1+2+3-4} )}{\Gamma (\Delta _1+2\Delta _2+is_{-2-3} )\Gamma (\Delta _2+is_{1 \pm 2} )\Gamma (\Delta _2+is_{\pm 3-4} )\Gamma (\Delta _1+is_{\pm 1-4} )\Gamma (\Delta _1+is_{3-2} )} + \\
& + \frac{\Gamma (\Delta _1+\Delta _2+is_{\pm 1-3} )\Gamma (\Delta _1+\Delta _2+is_{\pm 4-2} )\Gamma (2\Delta _2)\Gamma (is_{2+4-1-3} )\Gamma (is_{2-1-3-4} )\Gamma (2is_2)}{\Gamma (\Delta _1+2\Delta _2+is_{-2-3} )\Gamma (\Delta _2+is_{2 \pm 1} )\Gamma (\Delta _2+is _{\pm 4-3} )\Gamma (\Delta _1+is_{-1 \pm 4} )\Gamma (\Delta _1+is_{2-3} )} .
\end{split}
\end{equation}

\bibliography{main}
\bibliographystyle{JHEP}
\end{document}